\documentclass[11pt]{article}  

\usepackage{amsmath,amsthm,latexsym,amssymb,amsfonts,epsfig,braket,mathtools,yhmath}

\usepackage{cleveref,float}

\usepackage{subcaption}

\usepackage{xcolor}

\newcommand{\be}{\begin{equation}}
\newcommand{\ee}{\end{equation}}
\newcommand{\ba}{\begin{eqnarray}}
\newcommand{\ea}{\end{eqnarray}}
\newcommand{\bal}{\begin{align}}
\newcommand{\eal}{\end{align}}

\title{{\sf Hamiltonian renormalisation VIII. $P(\Phi)_2$ quantum field theory}}
\author{
{\sf M. Rodriguez Zarate}$^1$\thanks{{\sf 
melissa.rodriguez@gravity.fau.de}},
{\sf T. Thiemann}$^1$\thanks{{\sf 
thomas.thiemann@gravity.fau.de}}\\
\\
{\sf $^1$ Inst. for Theor. Phys. III, FAU Erlangen -- N\"urnberg,}\\
{\sf Staudtstr. 7, 91058 Erlangen, Germany}\\
}
\date{{\small\sf \today}}

\makeatletter
\@addtoreset{equation}{section}
\makeatother

\begin{document} 

\maketitle

{\sf

\begin{abstract}
In previous works in this series we focussed on Hamiltonian renormalisation of free 
field theories in all spacetime dimensions. In this paper we address the Hamiltonian renormalisation 
of the self-interacting scalar field in two spacetime dimensions with polynomial 
potential, called $P(\Phi)_2$. We consider only the finite volume case.

The $P(\Phi)_2$ theory is one of the few interacting QFT's that can be rigorously constructed 
non-perturbatively. We find that our Hamiltonian renormalisation flow finds this theory 
indeed as a fixed point. 
\end{abstract}

\section{Introduction}
\label{s1}

Constructing interacting quantum field theories (QFTs) rigorously in four and higher spacetime 
dimensions remains one of the most difficult challenges in theoretical and mathematical 
physics \cite{a}. The difficulties come from the fact that quantum fields are 
operator valued distributions which means that products thereof as they appear 
typically in Hamiltonians are a priori ill-defined, being plagued by both short distance 
(UV) and large distance (IR) divergences. In the constructive QFT (CQFT) approach \cite{b} 
one tames both types of divergences by introducing both UV ($M$) and IR cut-offs ($R$) to the effect 
that only a finite number of degrees survive at finite $M,R$. For instance, $R$ could be 
a compactification radius and $M$ a lattice spacing. Then at finite $M,R$ one is in the safe realm 
of quantum mechanics. The problem is then how to remove the cut-offs. Usually one removes 
first $M$ (continuum limit) and then $R$ (thermodynamic limit). In this process the parameters 
(coupling constants) are taken to be cut-off dependent and they are tuned or renormalised in such a way 
that the limiting theory is well-defined when possible.  

Non-perturbative renormalisation in CQFT (not to be confused with renormalisation 
in the perturbative approach to QFT) has a long tradition \cite{c} and comes 
in both the functional integral language and the Hamiltonian language (see e.g. \cite{d} and 
references therein). Focussing on UV cut-off removal, we consider quantum mechanical
systems labelled by the cut-off $M$. If these quantum mechanical systems all descend form 
a well-defined continuum theory, then in the functional integral approach one obtains 
the theory at resolution $M$ by integrating out all degrees of freedom referring to higher 
resolution while in the Hamiltonian approach one projects those out. This in particular implies 
that if one takes the quantum mechanical theory at resolution $M'$ and 
integrates or projects out the degrees of freedom at resolutions between 
$M<M'$ and $M'$ one obtains the quantum mechanical theory at resolution $M$. 
Vice versa, when this necessary set of {\it consistency conditions} is met, this typically 
also is sufficient to define a continuum theory.  
 
The family of theories that one starts with, are constructed making various choices such 
as representations, factor orderings, discretisation errors, etc. and the afore mentioned 
consistency conditions are generically violated. However, one can define  
a sequence of such quantum mechanical theory families by defining a new theory at resolution 
$M$ by integrating/projecting out the degrees of freedom between $M$ and $M'(M)$ of the old 
theory at resolution $M'(M)>M$ where $M':\; M\mapsto M'(M)$ is a fixed function on the set of 
resolution scales. Such a process is called a block spin transformation or coarse graining 
operation which typically leads to a renormalisation of the coupling constants. At a fixed point 
of this renormalisation flow of theories the consistency condition is enforced by construction
and therefore fixed points qualify as continuum theories.

In previous parts of this series we have considered a Hamiltonian projection scheme
\cite{LLT1,d} which is motivated by the functional integral approach via 
Osterwalder-Schrader reconstruction. It was then applied to free QFT 
in Minkowski space \cite{LLT2, LLT3, LLT4, LT, TT} in any dimension and parametrised 
QFT \cite{TZ} in 2d which shares some features with the free bosonic string. 
In all those cases the fixed point of the flow could be computed and was shown 
to coincide with the known continuum theory.
In the present paper for the first time we expose the formalism to
interacting quantum field theory (QFT), specifically to the self-interacting scalar 
quantum field theory in two spacetime dimensions with polynomial potential, called 
$P(\Phi)_2$ theory \cite{e}. We focus on the UV part of the renormalisation and 
thus consider the cylinder spacetime $\mathbb{R}\times [0,R)$ with periodic 
boundary conditions where the circumference  $2\pi\;R$ of the circle is fixed. Once
the theory is constructed at finite $R$ the thermodynamic limit $R\to \infty$ is taken 
by the methods described in \cite{e} and references therein should one 
be interested on the spacetime $\mathbb{R}^2$.  \\
\\
This work is organised as follows:\\

In section \ref{s2} we introduce the classical and quantum $P(\Phi)_2$ model on the cylinder. 
Its quantum theory is well defined in the Fock representation selected 
by the free part of its Hamiltonian no matter what the polynomial degree of its potential is,
even if it is not bounded from below, 
as long as the potential is normal ordered with respect to the same Fock structure. An 
elementary proof of this astonishing fact unique to two spacetime dimensions is provided 
in appendix \ref{sb}.

In section \ref{s3} we introduce the Hamiltonian renormalisation of this model. 
We pick the Dirichlet kernel \cite{l} to define the block spin transformation mentioned above. 
An account of Dirichlet kernel renormalisation techniques, which are closely related
to wavelet theory \cite{m}, may be found in appendix \ref{sa}. The Dirichlet kernel 
is a compromise between position locality and momentum decay properties, in particular it is 
smooth and therefore has advantages over the often chosen Schwarz kernel. The Schwarz kernel 
has better position locality properties but it has distributional derivatives which are potentially 
problematic with respect to treating the necessarily occurring field derivatives.    

In section \ref{s4} we derive the flow of the model and show that its fixed point
at finite resolution coincides with blocking from the known continuum QFT reviewed in section
\ref{s2}.

In section \ref{s5} we summarise and conclude.

\section{Classical and quantum $P(\Phi)_2$ on the cylinder}
\label{s2}

In the first subsection we introduce the classical $P(\Phi)_2$ theory and in the 
second we show that its normal ordered self-interacting Hamiltonian is a densely defined, symmetric 
operator in the Fock representation selected by its free part.

\subsection{Classical $P(\Phi)_2$ on the cylinder}
\label{s2.1}

We consider the cylinder spacetime manifold $\mathbb{R}\times [0,R)$ where $R$ is
any finite, fixed circumference of the circle. For convenience we pass from dimensionful coordinates 
$(c\;s,y)$ on that spacetime to dimensionless coordinates $X=(t=\frac{c\;s}{R},x=\frac{y}{R})$
so that $x\in [0,1)$ with endpoints identified. Then the classical action
can be rewritten as 
\be \label{2.1}
S[\Phi]=\int_{\mathbb{R}}\; dt\; \int_0^1\; dx\; \{
\frac{1}{2}[[\dot{\Phi}]^2-[\Phi']^2-p^2 \Phi^2]-P(\Phi)\}(X)
=:\int_{\mathbb{R}\times [0,1)}\; d^2\; L(\Phi(x),\Phi'(x),\dot{\Phi}(X)),
\ee
where p is the mass, $\dot{(.)}=\partial_t(.),\;(.)'=\partial_x(.)$ and
\be \label{2.2}
P(\Phi)=\sum_{k=0}^N\; g_k \; \Phi^k,
\ee
is any finite polynomial in the fields, i.e. a finite 
linear combination of powers $\Phi^k,\; k=0,..,N$ with real-valued, dimensionless
coefficients $g_k$ called coupling constants. We assume that whatever the value 
of $g_2$ is, we have $p^2=[M\;R]^2>0$ where $M$ is the dimensional mass. This can always be 
achieved by redefining $g_2$ and is done in order  
to avoid the special treatment of zero modes. We take $\Phi$ dimension free and have dropped 
a constant pre-factor from the action that has the dimension of an action. The spacetime field 
$\Phi$ is subject to periodic boundary conditions $\Phi(t,0)=\Phi(t,1)$ for all $t\in\mathbb{R}$.
Therefore in all that follows we never have to worry about boundary terms when integrating 
by parts w.r.t. $x$.   

The Hamiltonian formulation is straightforward: The real valued 
time zero field $\phi(x):=\Phi(0,x)$ 
has the real valued conjugate momentum $\pi(x)=[\partial_t \Phi](0,x)$ which obey the canonical Poisson brackets
\be \label{2.3}
\{\phi(x),\phi(y)\}=\{\pi(x),\pi(y)\}=0,\;\;
\{\pi(x),\phi(y)\}=\delta(x,y),
\ee
where $\delta(x,y)$ is the periodic delta distribution (see e.g. appendix \ref{sa})
and the Hamiltonian is 
obtained by the Legendre transformation 
\be \label{2.4}
h[\phi,\pi]={\sf extr}_u\;\{\int_0^1\; dx\; [\pi(x)\;u(x)-L(\phi(x),\phi'(x),u(x))]\}
=
\int_0^1\; dx\; \{\frac{1}{2}[\pi^2+[\phi']^2+p^2 \phi^2]+P(\phi)\}(x).
\ee

\subsection{Quantum $P(\Phi)_2$ on the cylinder}
\label{s2.2}

The structure of the Hamiltonian (\ref{2.4}) suggests the natural split
\be \label{2.5}
h=h_0+v,\;\;
h_0=
\int_0^1\; dx\; \{\frac{1}{2}[\pi^2+\phi\;\omega^2\cdot \phi\}(x),\;\;
v=\int_0^1\; dx\; P(\phi(x)),
\ee
where 
\be \label{2.5a}
\omega^2:=p^2-\Delta,\;\:\Delta=\partial_x^2
\ee
is a self-adjoint operator on the one particle Hilbert space
\be \label{2.6a}
L=L_2([0,1),\;dx),
\ee
with pure point spectrum $\hat{\omega}_n^2=p^2+[2\pi i\;n]^2,\; n\in \mathbb{Z}$. The corresponding 
eigenfunctions are $e_n(x)=e^{2\pi\; i\; n\;x}$ which define an orthonormal basis 
of $L$. For obvious reasons, $h_0,v$ are called the free part and interacting part respectively.

We define the unital Weyl $^\ast-$algebra $\mathfrak{A}$ in the usual way via its generating Weyl elements 
$w[f]=\exp(i<f,\phi>_L),\;w[g]=\exp(i<g,\pi>_L)$ for real valued $f,g\in L$ that are 
subject to the Weyl relations 
\be \label{2.6}
w[g]\;w[f]\;w[-g]=e^{-i\;<g,f>_L}\; w[f],\; w[f]\;w[f']=w[f+f'],\;w[g]\;w[g']=w[g+g'],\;
w[f]^\ast=w[-f],\;w[g]^\ast=w[-g].
\ee
We define a cyclic Fock representation $(\rho, {\cal H},\Omega)$ of $\mathfrak{A}$ from the Fock state\footnote{Note that the symbol $\omega$ is reserved for both the covariance operator and the Fock state introduced in \cref{2.7}; to distinguish the latter, it is denoted with a subscript $F$. In contrast, the letters $w$ and $W$ are used exclusively to denote Weyl elements.}
$\omega_F$ 
\be \label{2.7}
\omega_F(w[f]\;w[g]):=e^{\frac{i}{2}\;<f,g>_L}\;e^{-\frac{1}{4}[<f,\omega^{-1}\cdot f>_L+<g,\omega\cdot g>_L]}.
\ee
via the GNS construction. It is not difficult to show that (\ref{2.7}) is unitarily equivalent 
to defining annihilation operator valued distributions corresponding to the algebra element 
\be \label{2.8}
a(x):=\frac{1}{\sqrt{2}}[\omega^{1/2}\cdot\phi-i\omega^{-1/2}\cdot \pi](x),
\ee
so that $\rho(a(x))\Omega=0$ and everything else follows from the commutation relations of which 
the non-vanishing ones are
\be \label{2.8a}
[a(x),a^\ast(y)]=\delta(x,y).
\ee
The linear span $\cal D$ of the Fock vector states 
$\psi_{f_1,..,f_n}:=\rho(<f_1,a>^\ast)..\rho(<f_n,a>^\ast)\Omega$ for $f_1,..,f_n\in L$ 
together with $\Omega$ is dense in the Fock 
representation space $\cal H$. The number $n$ is called the particle number of the 
Fock vector where we assign $n=0$ to the vacuum $\Omega$ void of particles. We will 
in fact consider $\cal D$ corresponding to $f_1,..,f_n\in L_0$ where $L_0$ is 
the span of the functions $e_n$. This $\cal D$ is still dense and has the advantage 
that finite products of functions in $L_0$ are still in $L_0$.  

We use capital letters in order to denote the operator representatives of algebra elements,
e.g. $A(x)=\rho(a(x)),\;A^\dagger(x)=\rho(a^\ast(x))$. 
We use this Fock structure in order to define both $H_0$ and $V$ by their normal ordered 
symbols 
\ba \label{2.9}
H_0&:=& \rho(h_0)=\int_0^1\;dx\;A^\dagger(x) \;[\omega\cdot A](x),\;
V=\sum_{k=0}^N \;g_k V_k,\;
\nonumber\\
V_k &:=&\rho(v_k)=2^{-k/2}\;\sum_{l=0}^k\;
\left( \begin{array}{c} k \\ l \end{array} \right) \; V_k(l),\;
\nonumber\\
V_k(l) &:=& \int_0^1\; dx\;
\{[\omega^{-1/2}\cdot A^\dagger](x)\}^{k-l}\; [\omega^{-1/2}\cdot A](x)\}^l.\;
\ea
By construction, $\cal D$ is an invariant, dense domain of $H_0$. The astonishing fact,
unique to two spacetime dimensions, when $m>0$  and only when the spacetime is spatially compact, is that 
$\cal D$ is also a dense domain for $V$, albeit no longer an invariant one when $g_k\not=0$ 
for at least one of $k=2,..,N$. We give an elementary proof in appendix 
\ref{sb}.

\section{Hamiltonian renormalisation of $P(\Phi)_2$}
\label{s3}
An account of the version of Hamiltonian renormalisation used below and employing 
the Dirichlet kernel can be found in \cref{sa}, see \cite{d,LLT1} for its motivation.
Furthermore, in the following sections we adopt the notation introduced in \cref{sa}: capital letters such as \( F, G \) denote functions in \( L \); capital letters with a subscript \( M \), such as \( F_M, G_M \), refer to functions in \( L_M \); and lowercase letters with a subscript \( M \), such as \( f_M, g_M \), denote functions in \( l_M \). Elements of the abstract algebra are still written in lowercase, while their representatives (after GNS construction) are denoted by capital letters. \\

The starting point of the renormalisation scheme is to provide a family of theories 
$(\rho^{(0)}_M,\;{\cal H}^{(0)}_M,\;\Omega^{(0)}_M,\; H^{(0)}_M)$ consisting of a Hilbert space 
${\cal H}^{(0)}_M$ with cyclic 
vector $\Omega^{(0)}_M$ that carries a representation $\rho_M^{(0)}$ of some $^\ast-$algebra $\mathfrak{A}_M$ 
and a Hamiltonian operator $H^{(0)}_M$ densely defined on some subspace 
${\cal D}^{(0)}_M$ of ${\cal H}^{(0)}_M$. Equivalently, we consider a state $\omega^{(0)}_{M,F}$ on $\mathfrak{A}_M$
for which $(\rho^{(0)}_M,\;{\cal H}^{(0)}_M,\;\Omega^{(0)}_M)$ are its GNS data. The objects $\mathfrak{A}^{(0)}_M$ 
and $H^{(0)}_M$ are to be thought of as discretised versions of the continuum objects $\mathfrak{A}$ and 
$H$. The label $M$ is in general taken from a partially ordered and directed index set $\mathbb{O}$
that describes the (location, momentum, energy,..) resolution 
at which we probe the theory. In the present case the central tool is the Dirichlet kernel
\be \label{3.1}
P_M(x,y)=\sum_{n\in \mathbb{Z}_M}\; e_n(x-y),\;\; \mathbb{Z}_M=\{n\in \mathbb{Z};\;|n|\le \frac{M-1}{2}\},
\ee
which can be considered as a tamed version of the $\delta$ distribution on the circle. 
The index set is taken to be the set of odd naturals with ordering relation 
$M<M'$ iff $M'/M\in \mathbb{O}$. The Dirichlet kernel can be considered as an orthogonal 
projection $P_M$ in $L$ with image $L_M$. We define the Weyl algebra $\mathfrak{A}_M$ in 
analogy to (\ref{2.6}) as 
the abstract $^\ast-$algebra generated by the Weyl elements $w_M[F_M],\; w_M[G_M]$
with real valued $F_M,G_M\in L_M$
subject to the relations
\ba \label{3.2}
&&w_M[G_M]\;w_M[F_M]\; w_M[-G_M]=e^{-i<G_M,F_M>_{L_M}}\; w_M[F_M],\;\;
w_M[F_M]\;w_M[F'_M]=w_M[F_M+F'_M],\;
\nonumber\\
&& w_M[G_M]\;w_M[G'_M]=w_M[G_M+G'_M],\;
w_M[F_M]^\ast=w_M[-F_M],\; w_M[G_M]^\ast=w_M[-G_M],\;
\ea
where the scalar product on the finite resolution 1-particle Hilbert space $L_M$ coincides 
with the one on $L$.
The state $\omega^{(0)}_{M,F}$ is chosen to be in analogy to (\ref{2.7}) as
the Fock state 
\be \label{3.3}
\omega^{(0)}_{F,M}(w_M[F_M]\;w_M[G_M])=
e^{\frac{i}{2}\;<F_M,G_M>_{L_M}}\;e^{-\frac{1}{4}[<F_M,[\omega^{(0)}_M]^{-1}\cdot F_M >_{L_M}+<G_M,[\omega^{(0)}_M]\;\cdot G_M>_{L_M}]},
\ee
which requires as a central input the definition of the kernel $\omega^{(0)}_{M}$ on $L_M$. 
Note that the unfortunate doubling of symbols $\omega$ for both an algebraic state and 
a kernel on the 1-particle Hilbert space is resolved by attaching an extra index ``F'' 
to the state to indicate ``Fock''. To motivate the choice of $\omega^{(0)}_M$ we follow 
the general prescription of appendix \ref{sa} and define $h_M$ as a quantisation
of
\be \label{3.3a}
h_M[\phi_M,\pi_M]:=h[\phi_M,\pi_M],\; \phi_M=P_M\cdot \phi,\; \pi_M=P_M\cdot \pi,
\ee
where $h$ is the classical continuum Hamiltonian (\ref{2.5}). For its free part we find 
\begin{align} \label{3.4}
h_{0,M}&=\frac{1}{2}\;\int_0^1\; dx\; [\pi_M(x)^2+[\partial_x \phi_M(x)]^2+p^2\; \phi_M(x)^2]
\nonumber \\ & =\frac{1}{2} [<\pi_M,\pi_M>_{L_M}+<\partial\phi_M,\partial\phi_M>_{L_M}+\;p^2<\phi_M,\phi_M>_{L_M}].
\end{align}
Integrating the derivative term by parts (note that $P_M$ preserves the space of periodic 
$L_2$ functions) we obtain   
\be \label{3.6}
h_{0,M}=\frac{1}{2} [<\pi_M,\pi_M>_{L_M}-<\phi_M,\Delta_M\cdot\phi_M>_{L_M}+\;p^2\;<\phi_M,\phi_M>_{L_M}],
\ee
with the Laplacian
\be \label{3.77}
\Delta_M:=\partial_M^2,\; \partial_M=P_M\cdot \partial\cdot P_M.
\ee
We have used the projection property $P_M\cdot P_M=P_M$ of the Dirichlet kernel to 
obtain (\ref{3.77}). The specific form of $h_{0,M}$ suggests a Fock quantisation 
with annihilators 
\ba \label{3.7}
a_M(x)=2^{-1/2}[[\omega_M]^{1/2}\cdot \phi_M-i \; [\omega_M]^{-1/2}\cdot \pi_M](x),\;\;\;
[\omega_M]^2=p^2\; 1_M-\Delta_M,
\ea
which obey non-trivial commutation relations 
\be \label{3.8}
[a_M(x),a_M(y)^\ast]=P_M(x,y),
\ee
if one defines the Poisson brackets between $\phi_M, \pi_M$ as the result of considering 
these as functions on the continuum phase space and using the continuum Poisson bracket
\be \label{3.9}
\{\pi_M(x),\phi_M(y)\}:=\int\; du\; \int dv P_M(x,u)\; P_M(y,v)\;\{\pi(u),\phi(v)\}=P_M(x,y).
\ee
Then we see that (\ref{3.3}) and (\ref{3.7}) match provided that we interpret 
$w_M[F_M]=e^{i<F_M,\phi_M>_{L_M}}, \; w_M[G_M]=e^{i<G_M,\pi_M>_{L_M}}$. 

We can finish the initialisation of the Hamiltonian renormalisation flow 
by defining
\be \label{3.10}
H^{(0)}_M=H^{(0)}_{0,M}+V^{(0)}_M,\;
H^{(0)}_{0,M}=\int_0^1\;dx\; A_M^\dagger(x)\;[\omega^{(0)}_M\cdot A_M](x),\;
V_M=\int_0^1\; dx\; :V([[2\omega^{(0)}_M]^{-1/2}\cdot [A_M+A_M^\dagger]](x)):_M,
\ee
where $A_M$ is the operator representative of $a_M$ and $:.:_M$ denotes normal 
ordering of the $A_M, A_M^\dagger$.

\section{Hamiltonian renormalisation flow of $P(\Phi)_2$}
\label{s4}

In the first subsection we block the theory from the continuum in order 
to determine which fixed point family the renormalisation flow should find. In the 
second subsection, we use projector maps $P_M$ -as defined in \cref{sa}-, to project from $L$ to the $L_M$ subspaces and then 
compute the renormalisation flow. In the last subsection we discretise the fields, worked in the $l_M$ spaces of square summable 
sequences and compute the renormalisation flow. In the two frameworks we show that the renormalisation flow of the initial family 
defined in the previous section is already at its fixed point. The reason for displaying the strictly equivalent flows in terms 
of projections and discretisations respectively is that the former is in the spirit of renormalisation schemes outside a lattice context
while the former emphasises the traditional real space block spin interpretation of the renormalisation flow.   

\subsection{Blocking from the continuum}
\label{s4.1}

Blocking from the continuum means to define a state $\omega_{F,M}$ and a Hamiltonian 
$H_M$ out of the continuum state $\omega_F$ and Hamiltonian $H$ defined in section 
\ref{s2} via the formulas 
\begin{align}
\label{4.1}
\omega_{F,M}(w_M[F_M]\; w_M[G_M])&:=\omega_F(w[F_M]\;w[G_M]),\nonumber \\
<W_M[F_M]\Omega_M,\; H_MW_M[G_M]>_{{\cal H}_M}&:=<W[F_M]\Omega,\; H\; W[G_M]\Omega>_{{\cal H}}
\end{align}
where $(\rho_M, {\cal H}_M,\Omega_M)$ and $(\rho, {\cal H},\Omega)$ are the GNS data
of $\omega_{F,M}$ and $\omega_F$ respectively and we have denoted the operator 
representatives by capital letters, i.e. $W_M[F_M]=\rho_M(w_M[F_M])$ and 
$W[F]=\rho(w[F])$. On the right hand side of (\ref{4.1}) the elements $F_M, G_M \in L_M$
are to be considered as elements of the continuum 1-particle Hilbert space $L$ by trivial embedding $F_M\mapsto F$. 

We start with the first line of (\ref{4.1}) and use (\ref{2.7})
\be \label{4.2}
\omega_F(w[F_M]\;w[G_M]):=e^{\frac{i}{2}\;<F_M,G_M>_L}\;e^{-\frac{1}{4}[<F_M,\omega^{-1}\cdot F_M>_L+<G_M,\omega\cdot G_M>_L]}.
\ee
We have explicitly with $\hat{\omega}_n=\sqrt{p^2+[2\pi n]^2}$ using a resolution of 
identity with respect to the ONB $e_n$ of $L$
\ba \label{4.3}
&& <F_M,\omega^{-1} \cdot F'_M>_L
=\sum_{n\in \mathbb{Z}}\; <F_M,\omega^{-1}\;e_n>_L\; <e_n, F'_M>_L     
=\sum_{n\in \mathbb{Z}}\; \hat{\omega}_n^{-1}\;  <F_M,e_n>_L\; <e_n, F'_M>_L   \nonumber\\
&=& \sum_{n\in \mathbb{Z}_M}\; \hat{\omega}_n^{-1}\;  <F_M,e_n>_{L_M}\; <e_n, F'_M>_{L_M}
=:<F_M, \omega_M^{-1}\cdot F'_M>_{L_M},
\ea
where we used that $F_M, F'_M$ are orthogonal to the $e_n,\; n\not\in \mathbb{Z}_M$. To 
interpret the operator $\omega_M$ on $L_M$ we explicitly compute 
\ba \label{4.4}
&& \partial_M \cdot F_M
=\sum_{n\in \mathbb{Z}_M}\; e_n\; <e_n,\partial_M\cdot F_M>_{L_M}
=\sum_{n\in \mathbb{Z}_M}\; e_n\; <e_n,\partial\cdot F_M>_{L_M}
\nonumber\\
&=& -\sum_{n\in \mathbb{Z}_M}\; e_n\; <\partial\cdot e_n,F_M>_{L_M}
=\sum_{n\in \mathbb{Z}_M}\; [2\pi i n]\;e_n\; <e_n,F_M>_{L_M},
\ea
hence 
\be \label{4.5}
[p^2-\Delta_M] \cdot F_M
=\sum_{n\in \mathbb{Z}_M}\; \hat{\omega}_n^2\;e_n\; <e_n,F_M>_{L_M}.
\ee
It follows that $\omega_M=\sqrt{p^2-\Delta_M}$ which is in fact identical 
to the natural choice $\omega^{(0)}_M$ of section \ref{s3} used to define 
the initial family of states. Concluding we find that $\omega_{F,M}$ coincides with 
$\omega^{(0)}_{F,M}$ defined in (\ref{3.3}).

Considering the second line of (\ref{4.1}) we use the elementary identity 
\ba \label{4.6}
&& A(x)\; W[F]=W[F]\; W[F]^{-1}\; A(x)\; W[F]
=W[F]\;(A(x)-i [<[2\omega]^{-1/2}\cdot F, A+A^\dagger>_L, A(x)])
\nonumber\\
&=& W[F]\;(A(x)+i ([2\omega]^{-1/2}\cdot F)(x) 1_{{\cal H}_M}),
\ea
to find for the free part
\ba \label{4.7}
&& <W[F_M]\Omega,\; H_0\; W[F'_M]\Omega>_{{\cal H}}
=\int\; dx\; \int dy\; \omega(x,y)\; <A(x)\; W[F_M]\Omega,\; A(y)\; W[F'_M]\;\Omega>_{{\cal H}}
\nonumber\\
&=& \int\; dx\; \int dy\; \omega(x,y)\; ([2\omega]^{-1/2}\cdot F_M)(x)\; ([2\omega]^{-1/2}\cdot F'_M)(y)\;
<W[F_M]\Omega,\; W[F'_M]\;\Omega>_{{\cal H}} 
\nonumber\\
&=& \frac{1}{2}\; <\omega^{-1/2}\cdot F_M, \omega\cdot \omega^{-1/2}\cdot F'_M>_L \; <W_M[F_M]\Omega_M,\; W_M[F'_M]\Omega_M>_{{\cal H}_M}
\nonumber\\
&=& \frac{1}{2}\; <F_M, F'_M>_{L_M} \; <W_M[F_M]\Omega_M,\; W_M[F'_M]\Omega_M>_{{\cal H}_M},
\ea
where $\omega(x,y)=\sum_{n\in \mathbb{Z}}\; \hat{\omega}_n e_n(x-y)$ is the integral kernel 
of $\omega$ and in the step before the last one we used that we already know that the GNS data
of $\omega_{F,M}$ are those of $\omega^{(0)}_{F,M}$. Accordingly $H_M=H^{(0)}_M$ defined in (\ref{3.10}). 
For the interacting part it suffices to consider a field monomial of order $k$ 
\ba \label{4.8}
V_k &=& \int\;dx\; :\{([2\omega]^{-1/2}\cdot [A+A^\dagger])(x)\}^k:
=\sum_{l=0}^k\; 
\left( \begin{array}{c} k \\ l \end{array} \right) \; 2^{-k/2}\; V_k(l),\;
\nonumber\\
V_k(l) &=& \int\;dx\; 
([\omega^{-1/2}\cdot A]^\dagger(x)])^{k-l}\; 
(\omega^{-1/2}\cdot A](x)])^l.
\ea
Using (\ref{4.6}) it follows
\begin{align} 
\label{4.9}
&<W[F_M]\Omega,\; V_k(l)\; W[F'_M]\Omega>_{{\cal H}}
=i^{l-(k-l)} \; \int\; dx\; ([\omega^{-1/2} \cdot F_M](x))^{k-l}\;([\omega^{-1/2} \cdot F'_M](x))^l \;\times \nonumber \\
&<W[F_M]\Omega,\; W[F'_M]\;\Omega>_{{\cal H}}   \nonumber\\
&= i^{l-(k-l)} \; \int\; dx\; ([\omega_M^{-1/2} \cdot F_M](x))^{k-l}\;([\omega_M^{-1/2} \cdot F'_M](x))^l
<W_M[F_M]\Omega_M,\; W_M[F'_M]\;\Omega_M>_{{\cal H}_M}  
\nonumber\\
&=<W_M[F_M]\Omega_M,\; [V_k(l)]_M \;W_M[F'_M]\Omega_M>_{{\cal H}_M},
\end{align}
where 
\be \label{4.10}
[V_k(l)]_M=\int\;dx\; 
([\omega_M^{-1/2}\cdot A_M]^\dagger(x)])^{k-l}\; 
(\omega_M^{-1/2}\cdot A_M](x)])^l.
\ee
It follows that the initial family defined in section \ref{s3} coincides with 
the family blocked from the continuum. The reason for this is simply the identity
\be \label{4.11}
\omega\cdot P_M=P_M \cdot \omega_M,
\ee
which is due to the fact that the derivative operator $\partial$ preserves the subspace
$L_M$. This particular feature of the Dirichlet kernel is not shared by most other kernels
such as the Schwarz kernel and in that case the initial family of section \ref{s3} does not 
coincide with the family blocked from the continuum of section \ref{s4.1} even for 
free theories, see \cite{LLT1, LLT2, LLT3, LLT4}.

\subsection{Renormalisation flow in terms of the projected fields}
\label{s4.2}

By definition, for $M<M'$ we consider the trivial embedding $L_M\to L_{M'};\; F_M\mapsto F_{M'}$ 
which is possible because the $L_M$ are nested subspaces of $L$. Then the analog of (\ref{4.1}) 
is given by 
\begin{align} \label{4.12}
\omega^{(n+1)}_{F,M}(w_M[F_M]\; w_M[G_M])&:=\omega^{(n)}_{F,M'}(w_{M'}[F_M]\;w_{M'}[G_M]),\nonumber \\
<W_M[F_M]\Omega^{(n+1)}_M,\; H^{(n+1)}_M\; W_M[F'_M]\Omega^{(n+1)}_M>_{{\cal H}^{(n+1)}_M}
&:=
<W_{M'}[F_M]\Omega^{(n)}_{M'},\; H^{(n)}_{M'}\; W_{M'}[F'_M]\Omega^{(n)}_{M'}>_{{\cal H}^{(n)}_{M'}},
\end{align}
where $(\rho^{(n+1)}_M, {\cal H}^{(n+1)}_M,\Omega^{(n+1)}_M)$ and 
$(\rho^{(n)}_{M'}, {\cal H}^{(n)}_{M'},\Omega^{(n)}_{M'})$ are the GNS data
of $\omega^{(n+1)}_{F,M}$ and $\omega^{(n)}_{F,M'}$ respectively and we have denoted the operator 
representatives by capital letters, i.e. $W_M^{(n+1)}[F_M]=\rho^{(n+1)}_M(w_M[F_M])$ and 
$W_{M'}[F_{M'}]=\rho^{(n)}_{M'}(w_{M'}[F_{M'}])$. On the right hand side of (\ref{4.12}) the elements
$F_M, G_M, F'_M$ are to be considered as elements of the continuum 1-particle Hilbert space $L_{M'}$ by 
trivial embedding $F_M\mapsto F_M$. 

Going through literally the same steps as in section \ref{s4.1}, we find that the flow 
is trivial: $\omega^{(n)}_{F,M}=\omega^{(0)}_{F,M}=\omega^\ast_{F,M}=\omega_{F,M}$ and 
$H^{(n)}_M=H^{(0)}_M=H^\ast_M=H_M$, that is, every sequence element of the family coincides with 
the initial element which is also the element blocked from the continuum and thus the fixed 
point element. This is again due to (for $M'>M$)
\be \label{4.13}
\omega_{M'}\cdot P_M=P_M \cdot \omega_M.
\ee

\subsection{Renormalisation flow in terms of discretised fields}
\label{s4.3}

In order to test the version of Hamiltonian renormalisation method described in this series 
of papers less trivially for the solvable and self-interacting $P(\Phi)_2$ theory, one therefore should  
use an initial family which deviates from the natural choice of section \ref{s3}. For example
instead of the position non-local momentum ONB $e_n$ of $L_M,\; n\in \mathbb{Z}_M$ one may 
use the position quasi-local ONB $\chi^M_m,\; m\in \mathbb{N}_M=\{0,1,..,M-1\}$ of $L_M$ constructed in 
appendix \ref{sa}, see (\ref{a.7}), (\ref{a.8}). Indeed we can alternatively write 
(\ref{4.10}) as 
\be \label{4.14}
[V_k(l)]_M=M^{-k} \; \sum_{m_1,..,m_k\in \mathbb{N}_M}\; g_{M;m_1,m_2,..,m_k}\;
\prod_{s=1}^{k-l}\;[\omega_M^{-1}\cdot A_M]^\ast(m_s)\;
\prod_{s=k-l+1}^k\;[\omega_M^{-1}\cdot A_M](m_s),\;
\ee
where the "coupling constant" is given by 
\ba \label{4.15}
&& g_{M;m_1,m_2,..,m_k}=\int_0^1\;dx\; \prod_{s=1}^k \; \chi^M_{m_s}(x)
=\sum_{|n_1|,..,|n_k|\le (M-1)/2}\; \delta_{n_1+..+n_k=0}\; \prod_{s=1}^k \; e^{2\pi i n_k x^M_{m_s}}
\nonumber\\
&=& \sum_{|n_1|,..,|n_{k-1}|,|n_1+..+n_{k-1}|\le (M-1)/2}\; \prod_{s=1}^{k-1} \; e^{2\pi i n_k (x^M_{m_s}-x^M_{m_k})}
\ea
with $x^M_m=m/M$. If it was not for the constraint $|n_1+..+n_{k-1}|\le (M-1)/2$ this would collapse to 
\be \label{4.16}
\prod_{s=1}^{k-1}\; \chi^M_{m_s}(x^M_{m_k})=M^{k-1}\prod_{s=1}^{k-1}\; \delta_{m_s,m_k}
\ee
and we have
\be \label{4.17}
[\omega_M^{-1/2}\cdot A_M](m)=<\chi^M_m,\omega_M^{-1/2}\cdot A_M>_{L_M}=<\chi^M_m, \omega^{-1/2} A>_{L_M},
\ee
because $\chi^M_m\in L_M$ and due to (\ref{4.11}). Then (\ref{4.14}) could be considered as 
the $k-$fold Riemann sum of the coupling constant times the displayed polynomial  
of discretised annihilation and creation operators as $\epsilon_M=x^M_{m+1}-x^M_m=M^{-1}$.

However, due to the constraint the coupling
constant (\ref{4.15}) is only quasi-local and it is for (\ref{4.15}) that the flow is at its fixed point 
and not for (\ref{4.16}).
But one could start from the naive guess (\ref{4.16})
\be \label{4.18}
g^{(0)}_{M;m_1,m_2,..,m_k}=M^{k-1} \prod_{s=1}^{k-1} \; \delta_{m_s, m_k},
\ee
and then run the renormalisation flow. Already after the first iterations step one finds  
that the form (\ref{4.18}) is not preserved and that $g^{(1)}_{M;m_1,m_2,..,m_k}$ 
takes a
more general form. Starting from this more general form one sees that (\ref{4.15}) is a 
fixed point of the corresponding renormalisation flow equation. 

In more detail, we make the general Ansatz
\be \label{4.19}
[V^{(r)}_k(l)]_M=M^{-k} \; \sum_{m_1,..,m_k\in \mathbb{N}_M}\; g^{(r)}_{M; m_1,m_2,..,m_k}\;
\prod_{s=1}^{k-l}\;[\omega_M^{-1}\cdot A_M]^\ast(m_s)\;
\prod_{s=k-l+1}^k\;[\omega_M^{-1}\cdot A_M](m_s),\;
\ee
for the r-th renormalisation step with initial condition (\ref{4.18}). Then due to 
$\omega_{3M}^{-1}\; I_{M3M}= I_{M3M}\;\omega_{3M}^{-1}$ sandwiching 
$[V^{(r)}_k(l)]_{3M}$ between states 
of the form $w_{3M}[I_{M3M} f_M]\Omega_{3M}$ and requiring this to be 
$[V^{(r+1)}_k(l)]_M$
we see that we obtain 
\be \label{4.19a}
[V^{(r+1)}_k(l)]_M=(3M)^{-k} \; \sum_{m'_1,..,m'_k\in \mathbb{N}_{3M}}\; g^{(r)}_{3M;m'_1,m'_2,..,m'_k}\;
\prod_{s=1}^{k-l}\;[I_{M3M}\omega_M^{-1}\cdot A_M]^\ast(m'_s)\;
\prod_{s=k-l+1}^k\;[I_{M3M}\omega_M^{-1}\cdot A_M](m'_s),\;
\ee 
which produces the flow equation
\be \label{4.20}
g^{(r+1)}_{M;m_1,..,m_k}=3^{-k}\sum_{m'_1,..,m'_k}\; [\prod_{s=1}^k I_{M3M}(m'_s,m_s)]\;
g^{(r)}_{3M;m'_1,..,m'_k}
\ee
It is easy to see that (\ref{4.20}) has (\ref{4.15}) as a fixed point because 
\be \label{4.21}
\sum_{m'} I_{M3M}(m',m)\; \chi^{3M}_{m'}(x)
=M^{-1}\sum_{m'} <\chi^{3M}_{m'},\chi^M_m>\;\chi^{3M}_{m'}(x)
=3 (P_{3M}\chi^M_m)(x)=3\chi^M_m(x)
\ee
where completeness $\sum_m\; \chi^M_m(x)\chi^M_m(y)=M\;P_M(x,y)$ and $\chi^M_m\in L_M\subset L_{3M}$ 
was used. \\
\\
We now use (\ref{4.16}) as initial condition for the flow defined by (\ref{4.20}). We
need for $m'\in \mathbb{N}_{3M},\; m\in \mathbb{N}_M,\; n'\in \mathbb{Z}_{3M},\;
n\in \mathbb{Z}_M$
\be \label{4.22}
<\chi^{3M}_{m'},\chi^M_m>=\sum_{n,n'}\; e^{2\pi i\;[n'\;x^{3M}_{m'}-n\;x^M_m]}\;
<e^{2\pi n'\cdot },e^{2\pi i n\cdot}=\sum_n\; e^{2\pi i\;n[x^{3M}_{m'}-n\;x^M_m]}
\ee
since $\mathbb{Z}_M\subset \mathbb{Z}_{M'}$. Thus 
explicitly 
\ba \label{4.23}
&& g^{(1)}_{M;m_1,..,m_k}=\frac{1}{(3M)^k}\; \sum_{m'\in \mathbb{N}_{3M}^k} \; \prod_{s=1}^k\;
<\chi^{3M}_{m'_s},\chi^M_{m_s}> \; g^{(0)}_{3M;m'_1,..,m'_k}
\nonumber\\
&=& \frac{1}{3M}\; \sum_{m'\in \mathbb{N}_{3M}} \; \prod_{s=1}^k\;
<\chi^{3M}_{m'},\chi^M_{m_s}> 
\nonumber\\
&=& \frac{1}{3M}\; \sum_{n\in \mathbb{Z}_M^k}\;\sum_{m'\in \mathbb{N}_{3M}} \; 
\prod_{s=1}^k\;e^{2\pi i\;n_s[x^{3M}_{m'}-x^M_{m_s}]}
\nonumber\\
&=& \sum_{n\in \mathbb{Z}_M^k} \; 
[\prod_{s=1}^k\;e^{-2\pi i\;n_s\;x^M_{m_s}]}]\; \delta_{n_1+..+n_k,0\; ({\sf mod}\;3M)}
\ea
where in the last step we took care of the fact that summing over $m'\in \mathbb{N}_{3M'}$ 
produces $3M$ when $n_1+..+n_k$ is an integer multiple of $3M$ and zero else. Comparing 
with (\ref{4.15}) we see that the difference between the two expressions consists in the 
modulo $3M$ support of the Kronecker symbol because we can relabel $n_s\to -n_s$ as 
both sums are over the reflection invariant domain $\mathbb{Z}_M$.  

At the next iteration step we find 
\ba \label{4.24}
&& g^{(2)}_{M;m_1,..,m_k}=\frac{1}{(3M)^k}\; \sum_{m'\in \mathbb{N}_{3M}^k} \; \prod_{s=1}^k\;
<\chi^{3M}_{m'_s},\chi^M_{m_s}> \; g^{(1)}_{3M;m'_1,..,m'_k}
\nonumber\\
&=& \frac{1}{(3M)^k}\; \sum_{n'\in \mathbb{Z}_{3M}^k}\;\sum_{m'\in \mathbb{N}_{3M}^k} \; 
[\prod_{s=1}^k\;<\chi^{3M}_{m'_s},\chi^M_{m_s}> \; e^{-2\pi i\;n'_s\;x^{3M}_{m'_s}]}]\;
\delta_{n'_1+..+n'_k,0\; ({\sf mod}\;3 \; (3M))}
\nonumber\\
&=& \frac{1}{(3M)^k}\; \sum_{n'\in \mathbb{Z}_{3M}^k}\;\sum_{n\in \mathbb{Z}_M^k}\;\sum_{m'\in \mathbb{N}_{3M}^k} \; 
[\prod_{s=1}^k\;e^{-2\pi i n_s x^M_m}\;\; e^{2\pi i\;(n_s-n'_s)\;x^{3M}_{m'_s}]}]\;
\delta_{n'_1+..+n'_k,0\; ({\sf mod}\;3^2\;M)}
\nonumber\\
&=& \sum_{n'\in \mathbb{Z}_{3M}^k}\;\sum_{n\in \mathbb{Z}_M^k}\;
[\prod_{s=1}^k\;e^{-2\pi i n_s x^M_m}\;\delta_{n_s,n'_s}]\;
\delta_{n'_1+..+n'_k,0\; ({\sf mod}\;3^2\;M)}
\nonumber\\
&=& \sum_{n\in \mathbb{Z}_M^k}\;
[\prod_{s=1}^k\;e^{-2\pi i n_s x^M_m}]\;
\delta_{n_1+..+n_k,0\; ({\sf mod}\;3^2\;M)}
\ea
where we used that $\delta_{n'-n,0\;({\sf mod} 3M)}=\delta_{n'-n,0}$ as $|n'-n|\le 
\frac{3M-1+M-1}{2}<2M<3M$ i.e. the modulo operation can be dropped. We see that in the 
second iteration step the only change that happened compared to the first 
is that the modulo operation is now 
with respect to $3^2 M$ rather than $3^1 M$. 

Iterating (formally one proceeds by induction) we find 
explicitly for $r\ge 1$
\be \label{4.25}
g^{(r)}_{M;m_1,..,m_k}=\sum_{n\in \mathbb{Z}_M^k}\;
[\prod_{s=1}^k\;e^{-2\pi i n_s x^M_m}]\;
\delta_{n_1+..+n_k,0\; ({\sf mod}\;3^r\;M)}
\ee
Now $|n_1+..+n_k|\le k(M-1)/2$. For fixed resolution $M$, the number of steps $r$ required such that the modulo 
operation can be dropped is when that estimate is lower than the smallest possible nonzero 
integer multiple of $3^r M$ i.e. $k(M-1)/2<3^r\; M$ i.e. $3^r>k\frac{M-1}{2M}$. Thus independently 
of $M$ for a polynomial potential of degree $k$ we need at most $r_k:=1+[\frac{\ln(k/2)}{\ln(3)}]$ renormalisation 
steps until $g^{(r)}_{M;m_1,..,m_k}=g_{M;m_1,..,m_k}$ for all $r\ge r_k$ where $[.]$ denotes the Gauss bracket.
Note that this quick convergence of the flow would not occur had we not defined $\omega_M^2=p^2-\Delta_M,\;
\Delta_M=\partial_M^2,\; \partial_M=I_M^\dagger \partial I_M$ but instead had replaced $\partial_M$ by 
some of the more common discrete lattice derivatives (e.g. forward derivative 
$(\partial_M^+ f_M)(m)=M[f_M(m+1)-f_M(m)]$) because then the intertwining property 
$I_{M3M}\cdot \partial_M=\partial_{3M}\cdot I_{M3M}$ would not hold.\\
\\
We end this section by illustrating graphically the degree of non-locality of the fixed point 
coupling (\ref{4.15}) for the case $k=3$. Explicitly we consider 
\be \label{4.26}
f_M(y_1,y_2):=M^{-2}\sum_{|n_1|,|n_2|,|n_1+n_2|<(M-1)/2} \; e^{2\;i(n_1\; y_1+n_2\;y_2)}
\ee
which yields 
\be \label{4.27}
f(y_1=\pi [x^M_{m_1}-x^M_{m_3}],y_2=\pi [x^M_{m_2}-x^M_{m_3}])=M^{-2}\; g_{M;m_1,m_2,m_3} 
\ee
which we expect to be a quasi-local version of $\delta_{m_1,m_3}\;\delta_{m_2,m_3}$. The sums 
in (\ref{4.26}) can be computed in closed form using the geometric series summation formula
$\sum_{n=0}^{N-1}\;z^n=(z^N-1)/(z-1)$. We subdivide the summation domain into the three 
sets 
\ba \label{4.28}
&& \{n_1=\{1,..,\frac{M-1}{2}\},\;n_2=\{-\frac{M-1}{2},..,\frac{M-1}{2}-n_1\}\}
\nonumber\\
&\cup&
\{n_1=\{-1,..,-\frac{M-1}{2}\},\;n_2=\{-\frac{M-1}{2}-n_1,..,\frac{M-1}{2}\}\}  
\nonumber\\
&\cup &
\{n_1=\{0\},\;n_2=\{-\frac{M-1}{2},..,\frac{M-1}{2}\}\}
\ea
One finds  
\be \label{4.29}
f(y_1,y_2)=\frac{1}{2\;M^2}\{
\frac{\cos(M(y_1-y_2))}{\sin(y_1)\sin(y_2)}
+
\frac{\cos(M y_2)}{\sin(y_1)\sin(y_1-y_2)}
-
\frac{\cos(M y_1)}{\sin(y_2)\sin(y_1-y_2)}
\}
\ee
Despite its appearance, this function is everywhere smooth and bounded. 
Using $z_j=x_k-x_l,\;\epsilon_{jkl}=1$ with $y_1=x_1-x_3, y_2=x_2-x_3$ one can write this in the manifestly 
permutation invariant form 
\be \label{4.30}
f(z_1,z_2,z_3)=-\frac{1}{2\;M^2}
\{
\frac{\cos(M\;z_1)}{\sin(z_2)\sin(z_3)}
+
\frac{\cos(M\;z_2)}{\sin(z_3)\sin(z_1)}
+
\frac{\cos(M\;z_3)}{\sin(z_1)\sin(z_2)}
\}
\ee 
noticing the identity $z_1+z_2+z_3=0$ which displays the totally symmetric, smooth and 
bounded form (\ref{4.15})
of this function. For any $z_j=0\;\Rightarrow z_k=-z_l=z,\;\epsilon_{jkl}=1$ it reduces to 
\be \label{4.31}
\tilde{f}(z)=\frac{1}{2\;M^2\sin^2(z)}\;\{M\;\sin(M\;z)\sin(z)+1-\cos(M\;z)\cos(z)\}
\ee
as one can confirm by using de l'Hospital's theorem. Using de l'Hospital's theorem one 
more time when taking $z\to 0$ one can explicitly check that the maximum of this function is given not by unity but by 
$f(0,0)=\tilde{f}(0)=\frac{3}{4}+\frac{1}{4 M^2}$ corresponding to the fact that the constraints
$|n_1+n_2|\le (M-1)/2$ on $n_1,n_2$ roughly delete one quarter of the unconstrained $M^2$ points. Away from 
the maximum the function is of order $M^{-2}$ at a generic lattice point while it is 
of order $M^{-1}$ on one of the coordinate axes and on the diagonal $y_1=y_2$ (these are 
precisely the points $z_j=0,\; j=1,2,3$) but well away from the origin. This can be seen analytically 
noticing that $\cos(M\;z_j)=(-1)^{m_k-m_l},\;\epsilon_{jkl}=1$ while $\sin(M\;z_j)=0$. Accordingly 
if all $m_j-m_k$ are of the order $(M-1)/2$ the $\sin$ functions in the denominator of $f$ are close to unity.
If on the other hand say $m_2=m_3$ corresponding to $z_1=y_2=0, -z:=z_3=-z_2=\pi (m_1-m_2)/M=:m/M$
one finds $\tilde{f}(z)=\frac{1-(-1)^m \cos(m\pi/M)}{2\;M^2\sin^2(m\pi/M)}$ which for $m$ odd 
exceeds $ \frac{1}{2\;M^2\sin^2(m\pi/M)}$ and for large $M$ decreases quasi continuously from
approximatively  $[2\pi^2]^{-1}$ to $[2 M^2]^{-1}$ between $m=1$ and $m=(M-1)/2$. 
It is of order $\pi/(2M)$ for $m\approx\sqrt{M/\pi}$ which is well away from the vicinity of 
the origin but for large $M$ is outside of approximately the fraction $\sqrt{M}/M=1/\sqrt{M}$ 
of the available points.  \\
\\
In figures \ref{figure1}, \ref{figure2}, \ref{figure3} we illustrate this quasi-locality graphically for a few fixed resolutions 
$M=11,\;21,\;51,\;71,\; 111$
where we pick $m_3=5,\;10,\;25,\;35,\; 55$ 
respectively while 
$m_1,m_2 \in \mathbb{N}_M$ take full range. We computed the quasi-local coupling at the exact 
lattice points and then let Mathematica interpolate between those values.  
One sees the effect that the maximum $\approx 0.75$ becomes more 
pronounced and concentrated as the resolution increases while the function continues to display 
non-trivial oscillations in the vicinity of the maximum and along the coordinate axes and the diagonal.
\begin{figure}[H]
    \centering
    \begin{subfigure}{0.48\textwidth}
\includegraphics[width=\linewidth]{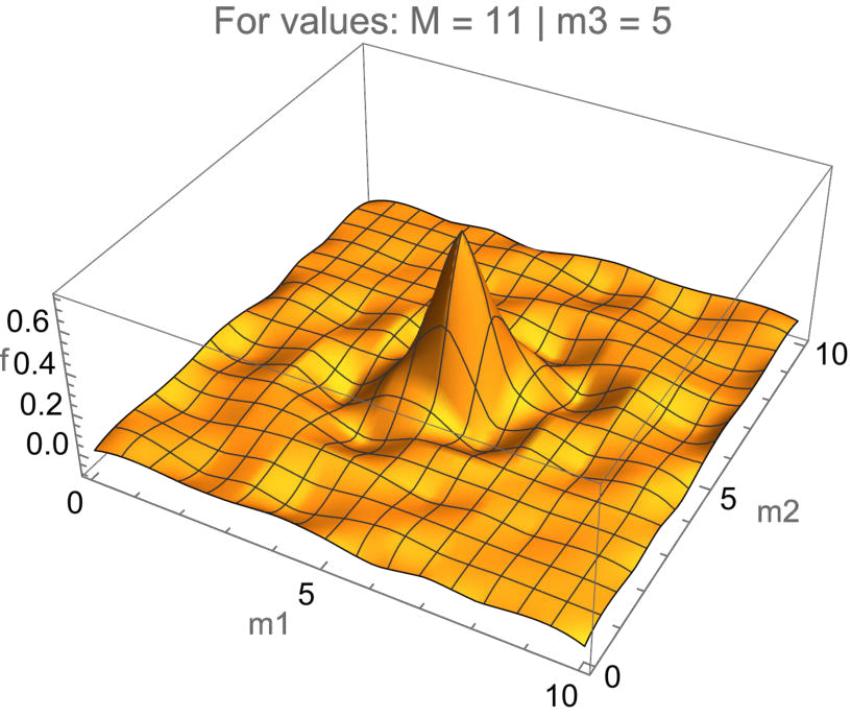}
\caption{}
    \end{subfigure}
    \hfill
    \begin{subfigure}{0.48\textwidth}
\includegraphics[width=\linewidth]{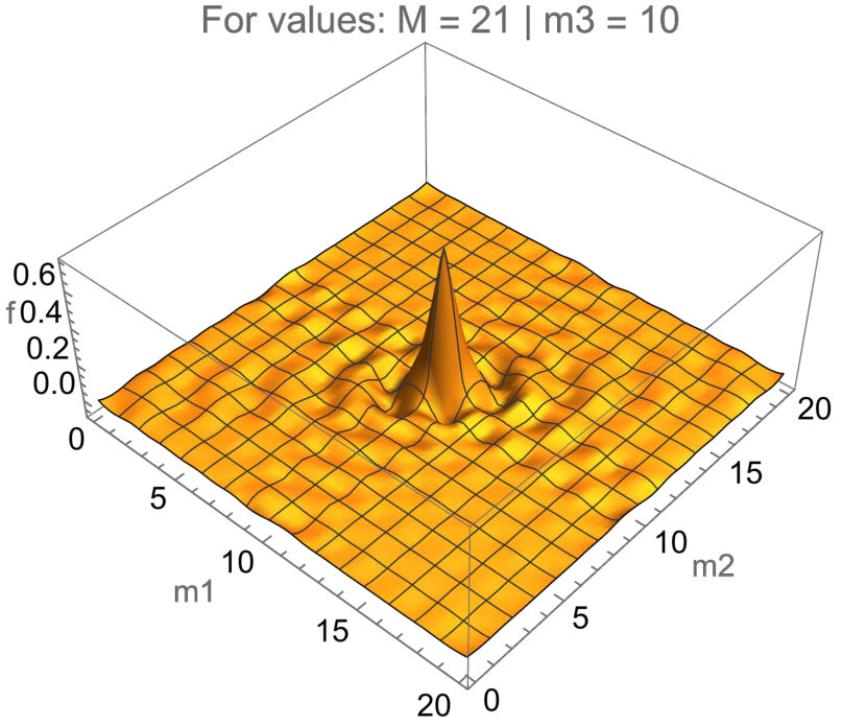}
    \end{subfigure}
     \caption{\textbf{Left:} Coupling for $M=11$ and $m_3=5$. \textbf{Right:} Coupling for $M=21$ and $m_3=10$.}
    \label{figure1}
\end{figure}
\begin{figure}[H]
    \centering
    \begin{subfigure}{0.48\textwidth}
\includegraphics[width=\linewidth]{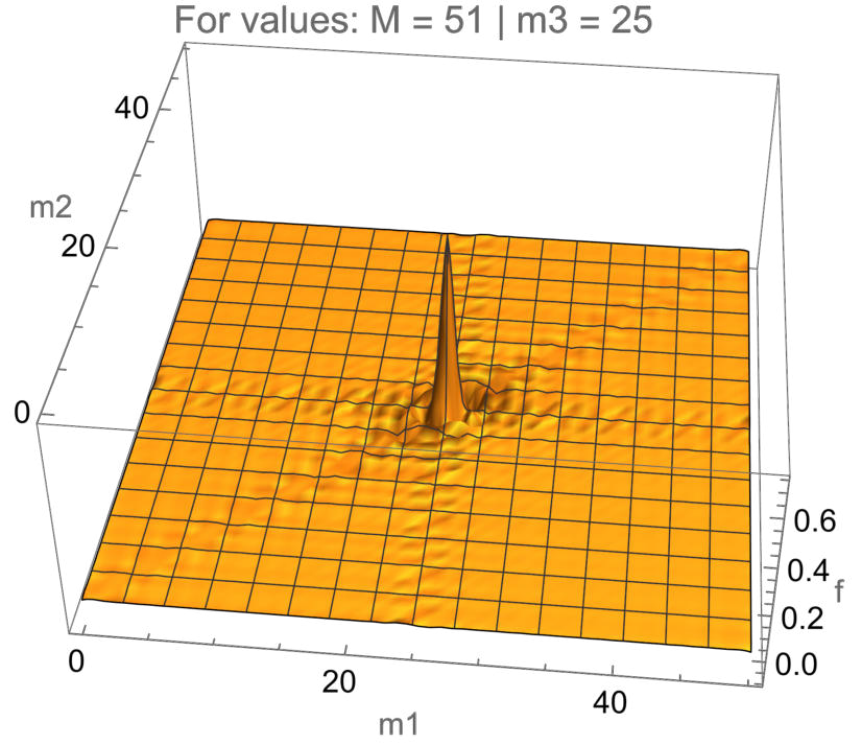}
\caption{}
    \end{subfigure}
    \hfill
    \begin{subfigure}{0.48\textwidth}
\includegraphics[width=\linewidth]{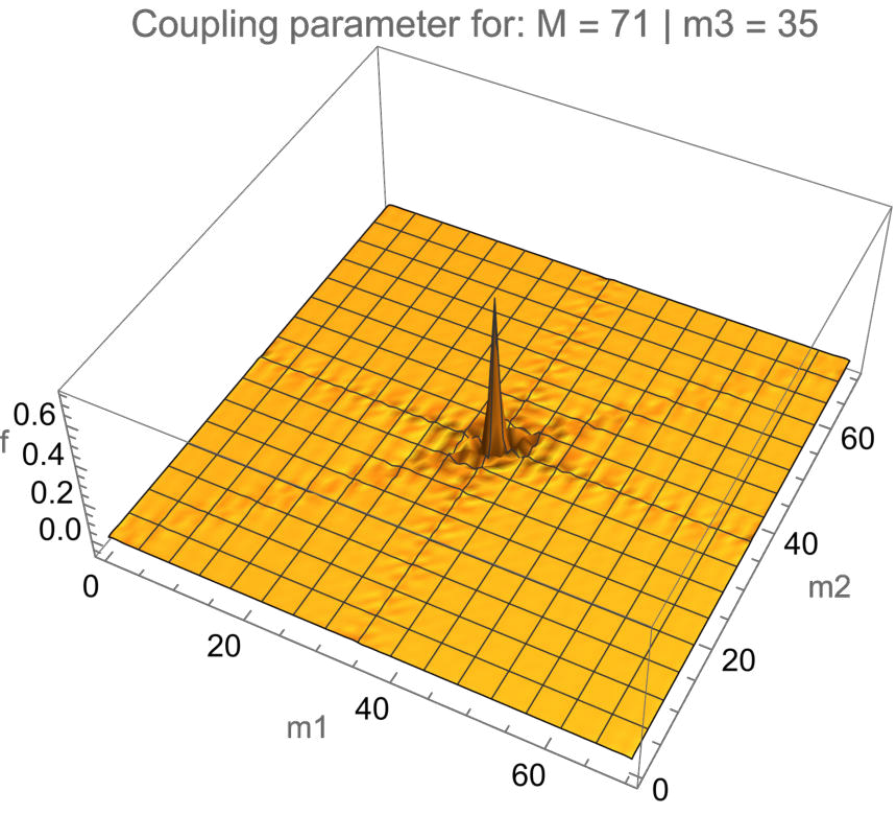}
    \end{subfigure}
     \caption{\textbf{Left:} Coupling for $M=51$ and $m_3=25$. \textbf{Right:} Coupling for $M=71$ and $m_3=35$.}
    \label{figure2}
\end{figure}
\begin{figure}[H]
    \centering
    \begin{subfigure}{0.48\textwidth}
\includegraphics[width=\linewidth]{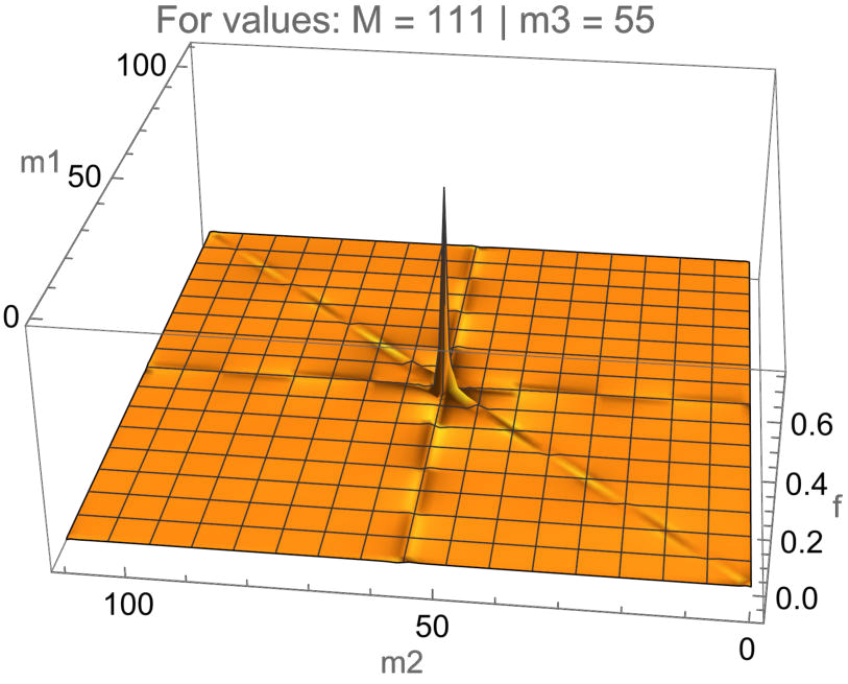}
\caption{}
    \end{subfigure}
    \hfill
    \begin{subfigure}{0.48\textwidth}
\includegraphics[width=\linewidth]{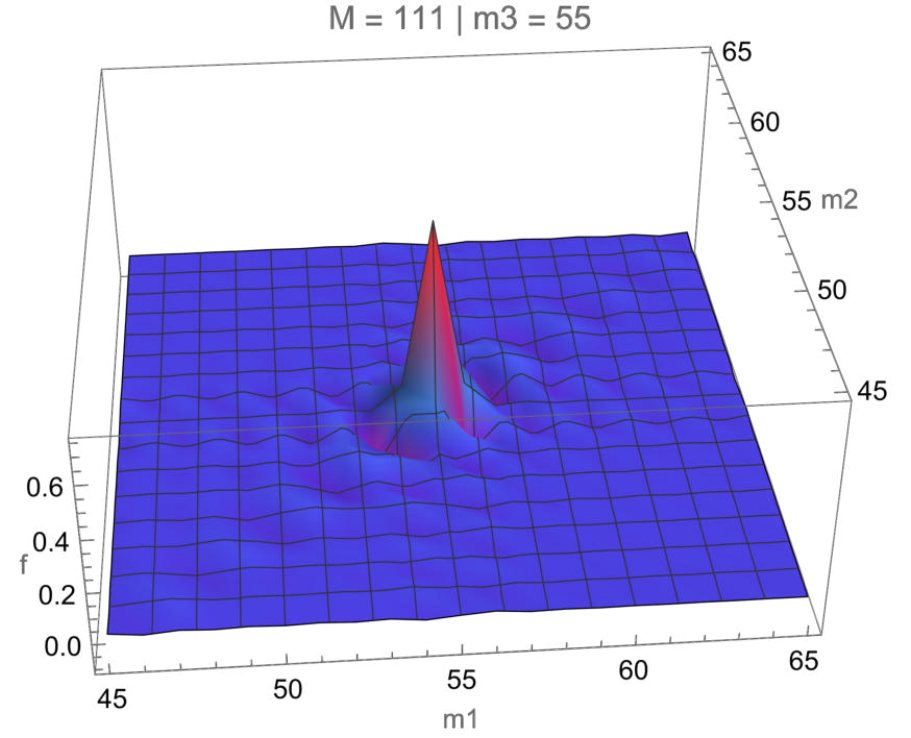}
    \end{subfigure}
     \caption{Coupling for $M=111$ and $m_3=55$. \textbf{Left:}  Full range. \textbf{Right:} Vicinity of maximum.}
    \label{figure3}
\end{figure}
%
%
%
In figure \ref{figure4} we zoom into the $\sqrt{M}$ vicinity of the maximum at $m_1=m_2=m_3=55$ for $M=111$. 
We cut off the maximum peak at a convenient value in order 
not to suppress the values of the coupling in its vicinity.  
\begin{figure}[H]
    \centering
    \begin{subfigure}{0.48\textwidth}
\includegraphics[width=\linewidth]{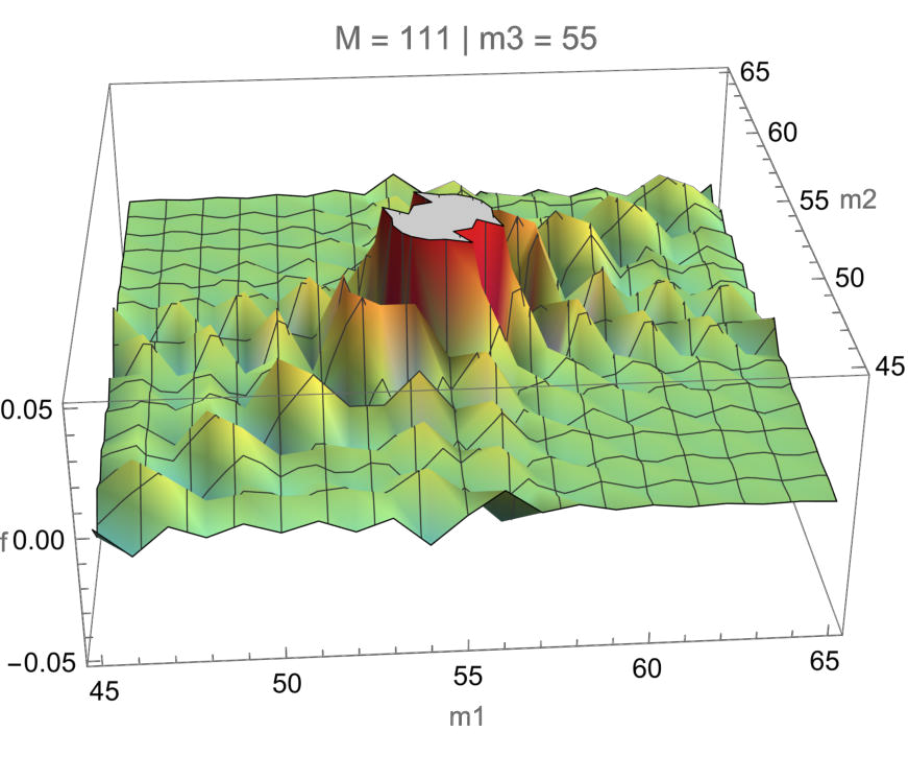}
\caption{}
    \end{subfigure}
    \hfill
    \begin{subfigure}{0.48\textwidth}
\includegraphics[width=\linewidth]{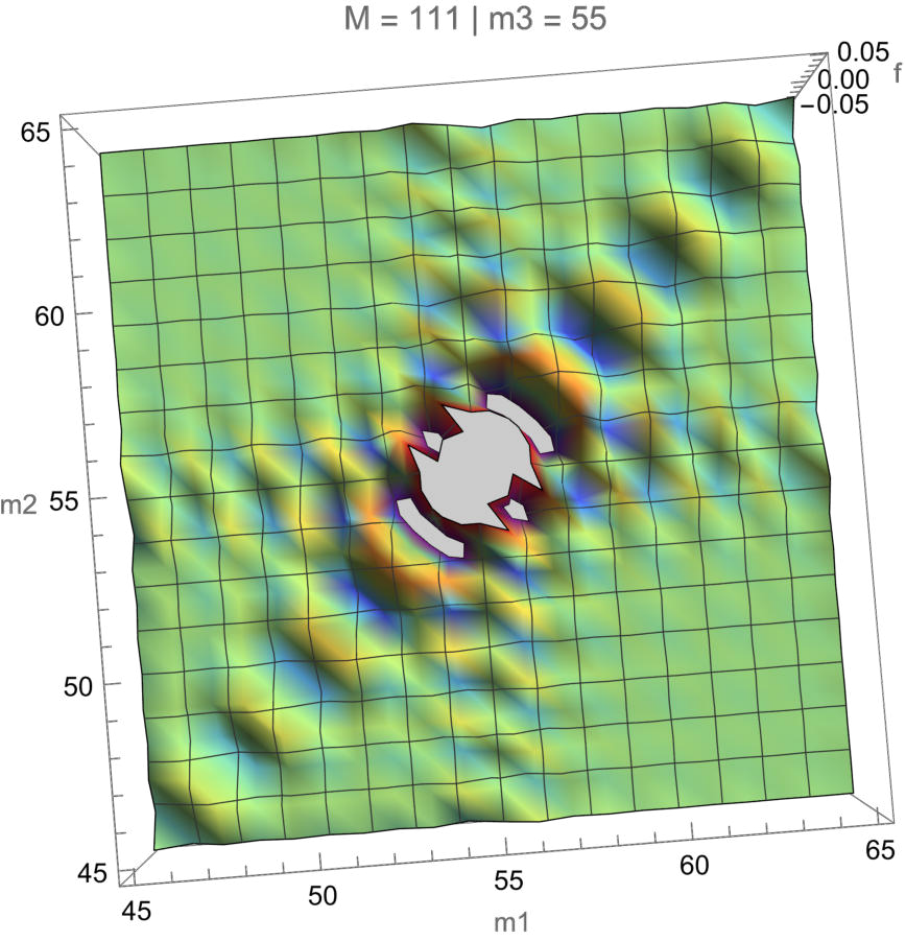}
    \end{subfigure}
     \caption{High resolution in the vicinity of the maximum for coupling for $M=111$ and $m_3=55$.\\ 
     \textbf{Left:} Side view.  \textbf{Right:} Top view.}
    \label{figure4}
\end{figure}

\section{Conclusions}
\label{s5}

We have successfully applied the version of Hamiltonian renormalisation developed in the 
current sequence of papers for the first time to an interacting QFT in finite volume in 
1+1 spacetime dimensions 
and showed that the renormalisation flow finds the known rigorous fixed point solution which is 
available in this case. The flow 
in fact stabilises rather quickly if one uses as coarse graining tools not the familiar 
position localised block spin transformations but rather smooth versions thereof which 
compromise between position and momentum locality and have better smoothness properties.
This demonstrates that the tools developed in the course of this series, which so far 
has focussed on free QFT, are also successfully applicable in the interacting case and 
does find the correct continuum Hamiltonian {\bf operator}.

Extrapolating to interacting QFT in higher, specifically four spacetime dimensions, the Hamiltonian 
$H=H_0+V$ when quantised in the Fock representation adapted to $H_0$ is known to be 
no longer an operator but merely a {\bf quadratic form}. Following the steps that 
we have carried out for the present model to any interacting QFT with Hamiltonian 
$H=H_0+V$ on the D+1 manifold $\mathbb{R}\times T^D$ shows that the corresponding flow 
would indeed find that quadratic form as a fixed point in the Fock representation adapted 
to $H_0$. However, ideally renormalisation
is designed to actually construct a continuum theory with $H$ as an operator rather than 
merely a quadratic form which by Haag's theorem \cite{a} requires to leave the realm 
of Fock representations adapted to $H_0$. To the best of our knowledge, the single 
known example where this actually worked in higher than 1+1 dimensions is $\Phi^4$ theory
in 2+1 dimensions \cite{k}. In addition to making the coupling constants $g_k$ to 
depend on $M$ (which in the present model was not necessary) a new and non-trivial 
step successfully performed in that seminal work was to invent invertible ``dressing transformations''
$T_M$ at resolution $M$ on the cut-off Fock space ${\cal H}_M$ with modes confined to 
$n\in \mathbb{Z}_M$ such that in $H'_M:=T_M^{-1}\; H_M T_M$ those terms are removed which prevent 
$H_M$ from being an operator rather than a quadratic form as $M\to \infty$ (these are 
normal ordered monomials in field operator valued distributions with more than one creation 
operator). If that is the case then ${\cal D}$ is an invariant domain for $H'_M$ where ${\cal D}$ is the 
span of Fock states in the free Fock space ${\cal H}$. Then 
$H_M$, defined on the new domain ${\cal D}'_M=T_M {\cal D}$ preserves ${\cal D}'_M$. 
The price to pay is that the norm of the vectors in ${\cal D}'_M$ with respect to the inner product 
$<.,.>$ on ${\cal H}$ diverges as $M\to\infty$. In \cite{k} it was shown that one can define 
a new inner product $<.,.>'_M:=\frac{<.,.>}{<T_M\Omega,T_M>}$ on ${\cal D}'_M$ 
where $\Omega$ is the Fock vacuum of $H_0$, such that one can take $M\to \infty$
and $H'_M\to H'$ becomes on operator densely defined on ${\cal D}'_M\to {\cal D}'$ whose 
Hilbert space completion with respect $<.,.>'_M\to <.,.>'$ results in a new Hilbert space 
${\cal H}'$ for the interacting theory. In order that this works, the norms of vectors $T_M \psi,\;
\psi\in {\cal D}$ with respect to $<.,.>$ must all diverge at the same rate. We expect 
that similar mechanisms must be invoked in the present renormalisation scheme as well.

In a forthcoming work we will apply the version of Hamiltonian renormalisation developed 
in this series of papers to the self-interacting $U(1)^3$ model of the weak Newton constant realm of Euclidian 
signature vacuum quantum gravity in four spacetime dimensions \cite{f}. This model has two known solutions, one 
in terms of operators \cite{g} using a Narnhofer-Thirring type of representation \cite{h}
and one in terms of quadratic forms \cite{i} in Fock representations \cite{i} in the sense 
that the full algebra of quantum constraints (Gauss, spatial diffeomorphism and Hamiltonian)
closes without anomaly. We will
investigate the Hamiltonian renormalisation flow for both types of representations with the aim
to gain new insights in how to renormalise full quantum gravity e.g. in the 
Loop Quantum Gravity (LQG) representation \cite{j} in order to remove or reduce present quantisation 
ambiguities.
\\ 
\\       
{\bf Acknowledgements}\\
M.R.-Z. thanks Jonas Neuser for useful discussions on the topic and acknowledges the financial support provided by the Deutsche 
Akademische Austauschdienst e. V.\\

\begin{appendix}

\section{Finiteness of $P(\Phi)_2$ QFT in Fock representations}
\label{sb} 

As motivated in section \ref{s2} we pick the Fock representation selected by 
the free part of the Hamiltonian $H_0$. We want to show that the interaction part
$V$ is densely defined on $\cal D$, the linear span of Fock vectors with smearing 
functions in $L_0$, the $L_2$ functions with compact momentum support. Below we 
present an elementary proof that requires just first year's calculus knowledge.\\
\\
To show this, consider a general Fock vector $\psi$. Then we have 
the elementary estimate 
\be \label{2.10}
||V\psi||_{{\cal H}}\le 
\frac{|g_k|}{2^{k/2}}\;\sum_{l=0}^k\;
\left( \begin{array}{c} k \\ l \end{array} \right) \; ||V_k(l)\psi||_{{\cal H}},\;
\ee
and writing $\psi=\sum_{n=0}^N \psi_n$ where $\psi_n$ is a Fock vector of particle number 
$n$ we have further the estimate
\be \label{2.11}
||V_k(l)\psi||_{{\cal H}}\le \sum_{n=l}^N\; ||V_k(l)\psi_n||_{{\cal H}}
\ee
Here we used that $V_k(l)$ annihilates $l$ particles and thus the r.h.s. of (\ref{2.11}) 
vanishes for $N<l$. Let then $n\ge l$ and $\psi_0=c \Omega,\;c=$const. or $\psi_n=
\prod_{r=1}\;[<f_r, A>]^\dagger\; \Omega,\; n>0$. Then for $l\ge 1$ we have
\be \label{2.12}
V_k(l)\psi_n=\sum_{I\in S_l}\;\int_0^1\; dx\; [[\omega^{-1/2}\cdot A^\ast](x)]^{k-l} \prod_{i\in I} [\omega^{-1/2}\cdot f_i](x)\;
\prod_{j\not\in I} <f_j, A>^\ast \; \Omega
\ee
where $S_l$ is the set of subsets of $\{1,..,n\}$ with $l$ elements  
and for $l=0$ we have 
\be \label{2.13}
V_k(0)\psi_n=\int_0^1\; dx\; [[\omega^{-1/2}\cdot A^\ast](x)]^k\;\psi_n
\ee
Using further elementary estimates it follows that $V$ is densely defined on $\cal D$ if and 
only if objects of the form 
\be \label{2.14}
\int_0^1\; dx\; F_l(x)\; [[\omega^{-1/2}\cdot A^\ast](x)]^{k-l}\; \psi_{n-l}
\ee
are normalisable vectors in the Fock space where $F_l$ is either the constant function   
equal to unity for $l=0$ or a product of $l$ functions in $L_0$ for $l>0$. It is understood 
that $k,n\ge l$ and $k=0,..,N$. For $k=0$ and thus $l=0$ there is nothing to show 
as (\ref{2.14}) just equals $<1,1>\;\psi_n$ which is finite as $1\in L$ when space is 
compact. For $k=1$ (\ref{2.14}) equals $<1,A>^\ast \psi_n$ 
and $<1,F_1>\;\psi_{n-1}$ respectively which are both Fock vectors. The interesting 
terms therefore come form $k\ge 2$.       

The norm squared of (\ref{2.14}) is 
\be \label{2.15}
\int_0^1\; dx\; \int_0^1\;dy\; F_l^\ast(x)\;F_l(y)\;
<\psi_{n-l},\;[[\omega^{-1/2}\cdot A](x)]^{k-l}\; 
  [[\omega^{-1/2}\cdot A^\ast](y)]^{k-l}\; \psi_{n-l}
\ee 
When moving the annihilators to the right we get a sum of terms with $0\le r \le k-l$ 
factors of 
\be \label{2.16}
[(\omega^{-1/2}\cdot A)(x)\;,\; 
(\omega^{-1/2}\cdot A^\ast)(y)]=\omega^{-1}(x,y)=\sum_{n\in\mathbb{Z}}\;\hat{\omega}_n^{-1}\; e_n(x-y)
\ee
which is the integral kernel of the operator $\omega^{-1}$ and another $k-l-r$ factors 
of functions in $L_0$ (for $k-l-r<0$ that term vanishes). Altogether we see that $V$ is 
densely defined on $\cal D$ if and only if integrals of the 
form 
\be \label{2.17}
\int_0^1\; dx\; \int_0^1\; dy\; F_{k-r}^\ast(x)\; G_{k-r}(y)\; [\omega{-1}(x,y)]^r
\ee
converge with $2\le k\le N$ and $0\le r\le k$ with $F_{k-r}, G_{k-r}\in L_0$ when 
$k-r>0$ and $F_0=G_0=1$. Let $\hat{F}_n=<e_n,F>$ be the Fourier transform 
of $F\in L$, then (\ref{2.17}) is either 
\be \label{2.18}
\sum_{K\in \mathbb{Z}}\; \hat{F}^\ast_{k-r}(K)\;\hat{G}_{k-r}(K)\;
\sum_{n_1,..,n_r\in \mathbb{Z}}\;\delta_{n_1+..+n_r,K} \;\prod_{s=1}^r \hat{\omega}_{n_s}^{-1}
\ee
for $k-r>0$ or 
\be \label{2.19}
\sum_{n_1,..,n_r\in \mathbb{Z}}\;\delta_{n_1+..+n_r,0} \;\prod_{s=1}^r \hat{\omega}_{n_s}^{-1}
\ee
for $k-r=0$ where again compactness of space was important as 
the kernel is translation invariant. As functions in $L_0$ have compact momentum support, 
the sum over $K$ in (\ref{2.18}) is finite and it is sufficient to show that the sums
\be \label{2.20}
\sum_{n_1,..,n_r\in \mathbb{Z}}\;\delta_{n_1+..+n_r,K} \;\prod_{s=1}^r \hat{\omega}_{n_s}
\ee
converge for every $2\le r\le N$ and any finite $K\in \mathbb{Z}$.  
 
The intuitive reason why this is in fact the case is as follows: If it was not 
for the Kronecker symbol, the sums would decouple into a product of $r$ sums 
each of which behaves as the $M\to\infty$ limit of $\zeta_M(1)$ which diverges
as $\ln(M)$ and where 
$\zeta_M(z)=\sum_{n=1}^M\; n^{-z}$ is the $M$ cut-off of the Riemann $\zeta$ function.
However due to the Kronecker symbol we only get $r-1$ sums over $n_1,..,n_{r-1}$ 
but still $r$ factors, one depending on $|n_s|,\;s=1,..,n_{r-1}$ and one depending 
on $|n_1+..+n_{r-1}-K|$ which roughly adds an additional factor of $\hat{\omega}_{n_s}^{1/(r-1)}$. 
Thus each of the $r-1$ sums behave as 
$\zeta_M(1+\frac{1}{r-1})$ which converges for every $r>1$. Of course this heuristic
argument is not rigorous.

We therefore rigorously estimate the multiple sum in (\ref{2.20}) as follows: First we simplify
\ba \label{2.21}
&&
\sum_{n_1,..,n_r\in \mathbb{Z}}\; \prod_{s=1}^r\; \hat{\omega}_{n_s}^{-1}\;
\delta_{n_1+..+n_r,K}
\le
\sum_{n_1,..,n_{r-1}\in \mathbb{Z}\cup \{0\}}\; \prod_{s=1}^r\; \hat{\omega}_{n_s}^{-1}\;
\delta_{n_1+..+n_r,K}
\\
&=& 
\sum_{\sigma_1,..,\sigma_r=\pm 1}\;
\sum_{n_1,..,n_r\in \mathbb{N}_0}\; \prod_{s=1}^r \;\hat{\omega}_{\sigma_s n_s}^{-1}\;
\delta_{\sigma_1 n_1+..+\sigma_r n_r,K}
\nonumber\\
&=& 
\sum_{\sigma_1,..,\sigma_r=\pm 1}\;
\sum_{n_1,..,n_r\in \mathbb{N}_0}\; \prod_{s=1}^r\; \hat{\omega}_{n_s}^{-1}\;
\delta_{\sigma_1 n_1+..+\sigma_r n_r,K}
\nonumber\\
&=& 
\sum_{t=0}^r\;
\left( \begin{array}{c} r \\ t \end{array} \right)\;
\sum_{n_1,..,n_r\in \mathbb{N}_0}\; \prod_{s=1}^r \hat{\omega}_{n_s}^{-1}\;
\times\nonumber\\
&& \{
\theta(K-1)\;\delta_{[n_1+..+n_t]-[n_{t+1}+..+ n_r+|K|],0}
+\theta(-K-1)\;\delta_{[n_1+..+n_t+|K|]-[n_{t+1}+..+ n_r],0}
+\delta_{K,0}\;\delta_{[n_1+..+n_t]-[n_{t+1}+..+ n_r],0}
\}
\nonumber
\ea
where we made use of reflection invariance $\hat{\omega}_n=\hat{\omega}_{-n}$ and that 
the product of the inverse eigenvalues is invariant under permutations of $r$ summation variables
$n_1,..,n_r$ so that the sum over $\sigma_1,.. \sigma_r$ can be reduced to 
a summation over the number $t=0,1,..,r$ of those $\sigma_1,.. \sigma_r$ which take 
the value $+$ which can be taken to be the $\sigma_1,..,\sigma_t$. 

The first term vanishes when $t=0$ and for $t=r$ the sum is constrained by $n_1,..,n_r\le |K|$  
as all summation variables are non negative. Likewise, the second term vanishes for $t=r$ and for 
$t=0$ the sum is constrained by $n_1,..,n_r\le |K|$. Finally the third term collapses 
to $n_1=..=n_r=0$ for $t=0, t=r$. Thus the terms with $t=0, r$ are trivially finite and 
it is sufficient to consider only the terms $0<t<r$. For those, in the second term we relabel 
$t\to r-t$ and switch the two groups $n_1,..,n_t$ and $n_{t+1},..,n_r$. As 
$\theta(K-1)+\theta(-K-1)+\delta_{K,0}=1$ we end up with the relevant contribution
\ba \label{2.22}
S_r(K) &:=& \sum_{t=1}^{r-1}\;
\left( \begin{array}{c} r \\ t \end{array} \right)\;
\sum_{n_1,..,n_r\in \mathbb{N}_0}\; \prod_{s=1}^r \hat{\omega}_{n_s}^{-1}\;
\delta_{[n_1+..+n_t]-[n_{t+1}+..+ n_r+|K|],0}    
\\
&=& m^r\;
\sum_{t=1}^{r-1}\;
\left( \begin{array}{c} r \\ t \end{array} \right)\;
\sum_{L=|K|}^\infty\; 
[\sum_{n_1,..,n_t\in \mathbb{N}_0} \; \prod_{s=1}^t \;w_{n_s}^{-1}\;\delta_{n_1+..+n_t,L}]\;
[\sum_{n_{t+1},..,n_r\in \mathbb{N}_0} \; \prod_{s=t+1}^r \;w_{n_s}^{-1}\;\delta_{n_{t+1}..+n_r,L-|K|}]\;
\nonumber
\ea
where $\kappa:=\frac{2\pi}{m}$ and $w_n=\sqrt{1+[\kappa n]^2}$ and $m$ is the mass parameter. 
Using the 
the abbreviation for $1\le t\le r-1,\; M\ge 0$ 
\be \label{2.23}
s_t(M)=\sum_{n_1,..,n_t\in \mathbb{N}_0} \; \prod_{s=1}^t \;w_{n_s}^{-1}\;\delta_{n_1+..+n_t,M}
\ee
which for $t=1$ just equals $w_{M}^{-1}$ we have 
\be \label{2.24}
S_r(K)=m^r\; \sum_{t=1}^{r-1}\;
\left( \begin{array}{c} r \\ t \end{array} \right)\;
\sum_{L=|K|}^\infty\; s_t(L)\; s_{r-t}(L-|K|)
\ee
The idea is now to show that $s_t(M)$ decays at least as $|M|^{-p}$ for some $p>1/2$ no matter 
the value of $t$ for then the sum over $L$ converges and the sum over $t$ is anyway finite. For 
$t=1$ this is trivial, hence in what follows we consider $t\ge 2$.

To this end, we use the basic estimate
\be \label{2.25}
1+x^2\ge c\; (1+|x|)^2\;\Leftrightarrow\; [|x|-\frac{c}{1-c}\;m]^2+\frac{1-2c}{(1-c)^2} m^2\ge 0
\ee
which holds for $0<c\le \frac{1}{2}$. We pick $c=\frac{1}{4}$ and thus
\be \label{2.26}
w_n^{-1}\le \frac{2}{1+\kappa n}
\ee
It follows
\ba\label{2.27}
&& s_t(M)\le 2^t\;\sum_{n_1=0}^M\; [1+\kappa n_1]^{-1}\; \sum_{n_2=0}^{M-n_1}\; [1+\kappa n_2]^{-1}\;
...
\\
&& \sum_{n_{t-2}=0}^{M-[n_1+..+n_{t-3}]}\; [1+\kappa n_{t-2}]^{-1}\;
\sum_{n_{t-1}=0}^{M-[n_1+..+n_{t-2}]}\; [1+\kappa n_{t-1}]^{-1}\;[1+\kappa (M-[n_1+..+n_{t-1}])]^{-1}
\nonumber
\ea
To estimate the sum over $n_{t-1}$ we use the abbreviation $l=M-[n_1+..+n_{t-2}]$ (which is defined 
to equal $M$ when $t=2$)
\ba\label{2.28}
&& \sum_{n=0}^l\; [1+\kappa n]^{-1}\;[1+\kappa (l-n)]^{-1}
=[1+\kappa\; l]^{-1} \sum_{n=0}^l\; ([1+\kappa n]^{-1}+[1+\kappa (l-n)]^{-1})
=\frac{2}{1+\kappa l}\;\sum_{n=0}^l\; [1+\kappa n]^{-1}
\nonumber\\
&=& \frac{2}{1+\kappa l}\;[1+\sum_{n=1}^l\; [1+\kappa n]^{-1}]
=\frac{2}{1+\kappa l}\;[1+\sum_{n=1}^l\;\int_{n-1}^n\; dk\; [1+\kappa n]^{-1}]
\le\frac{2}{1+\kappa l}\;[1+\sum_{n=1}^l\;\int_{n-1}^n\; dk\; [1+\kappa k]^{-1}]
\nonumber\\
&=& \frac{2}{1+\kappa l}\;[1+\int_0^l\; dk\; [1+\kappa k]^{-1}]
=\frac{2}{1+\kappa l}\;[1+\kappa^{-1}\;\ln(1+\kappa l)]
\ea
Consider now any $0<p<1$ and note that $z=:1+\kappa l\ge 1$ then 
\be \label{2.29}
\frac{\ln(z)}{z}\le z^{-p} (1-p)^{-1}, \; z^{-1}\le z^{-p}
\ee
Thus there exists a constant $c_1$ (depending on $p_1,\kappa$) 
such that (\ref{2.28}) is bounded from above 
by $c_1\;[1+\kappa l]^{- p_1}$ for any $0<p_1<1$. It follows that 
\ba \label{2.30}
&& s_t(M)\le 2^t\;c_1\;\sum_{n_1=0}^M\; [1+\kappa n_1]^{-1}\; \sum_{n_2=0}^{M-n_1}\; [1+\kappa n_2]^{-1}\;
...
\\
&& \sum_{n_{t-3}=0}^{M-[n_1+..+n_{t-4}]}\; [1+\kappa n_{t-3}]^{-1}\;
\sum_{n_{t-2}=0}^{M-[n_1+..+n_{t-3}]}\; [1+\kappa n_{t-2}]^{-1}\;[1+\kappa (M-[n_1+..+n_{t-2}])]^{-p_1}
\nonumber
\ea
To estimate the sum over $n_{t-2}$ we use the abbreviation $l=M-[n_1+..+n_{t-3}]$ (which is defined 
to equal $M$ when $t=3$) and use again $z^{-1}<z^{-p_1}$ for $z\ge 1$ to find 
\be \label{2.31}
\sum_{n=0}^l\; [1+\kappa n]^{-1}\;[1+\kappa (l-n)]^{-p_1}
\le \sum_{n=0}^l\; [1+\kappa n]^{-p_1}\;[1+\kappa (l-n)]^{-p_1} 
\le [1+\kappa l]^{-p_1}\; \sum_{n=0}^l\; [[1+\kappa n]^{-1}+[1+\kappa (l-n)]^{-1}]^{p_1} 
\ee
For any two numbers $a,b>0$ and $p>0$ we have 
\be \label{2.32}
[a+b]^p
=\left( \begin{array}{cc} 
a^p\; [1+b/a]^p & a\ge b \\ 
b^p\; [1+a/b]^p & b\ge a 
\end{array} \right.
\le \left( \begin{array}{cc} 
(2a)^p & a\ge b \\ 
(2b)^p\; & b\ge a 
\end{array} \right.
\le 2^p\;[a^p + b^p]
\ee
Therefore (\ref{2.31}) can be further estimated by 
\ba \label{2.33}
&& 2^{p_1}\; [1+\kappa l]^{-p_1}\; \sum_{n=0}^l\; \{[1+\kappa n]^{-p_1}+[1+\kappa (l-n)]^{-p_1}\}
=2^{1+p_1}\; [1+\kappa l]^{-p_1}\; \sum_{n=0}^l\; [1+\kappa n]^{-p_1}
\nonumber\\
&=& \frac{2^{1+p_1}}{[1+\kappa l]^{p_1}}\;[1+\sum_{n=1}^l\; [1+\kappa n]^{-p_1}]
=\frac{2^{1+p_1}}{[1+\kappa l]^{p_1}}\;[1+\sum_{n=1}^l\;\int_{n-1}^n\; dk\; [1+\kappa n]^{-p_1}]
\nonumber\\
&\le& \frac{2^{1+p_1}}{[1+\kappa l]^{p_1}}\;[1+\sum_{n=1}^l\;\int_{n-1}^n\; dk\; [1+\kappa k]^{-p_1}]
\nonumber\\
&=& \frac{2^{1+p_1}}{[1+\kappa l]^{p_1}}\;[1+\int_0^l\; dk\; [1+\kappa k]^{-p_1}]
=\frac{2^{1+p_1}}{[1+\kappa l]^{p_1}}\;[1+\kappa^{-1}\;(1-p_1)^{-1}\; ([1+\kappa l]^{1-p_1}-1)] 
\nonumber\\
&\le& 
\frac{2^{1+p_1}}{[1+\kappa l]^{p_1}}\;[|1-\frac{1}{\kappa (1-p_1)}|
+\kappa^{-1}\;(1-p_1)^{-1}\; [1+\kappa l]^{1-p_1}] 
\ea
We choose $p_1$ so close to $1$ that $p_2:=2p_1-1$ still obeys $0<p_2<1$ which is the 
case for $1/2<p_1<1$. Then $p_2<p_1$ so that also $[1+\kappa l]^{-p_1}<[1+\kappa l]^{-p_2}$. 
Thus there exists a constant $c_2$ (depending on $p_1,\kappa$) 
such that (\ref{2.33}) is bounded from above 
by $c_2\;[1+\kappa l]^{- p_2}$ for any $1/2<p_1<1$ with $p_2=2 p_1 -1$. 
It follows that 
\ba \label{2.34}
&& s_t(M)\le 2^t\;c_1\;c_2\;\sum_{n_1=0}^M\; [1+\kappa n_1]^{-1}\; \sum_{n_2=0}^{M-n_1}\; [1+\kappa n_2]^{-1}\;
...
\\
&& \sum_{n_{t-4}=0}^{M-[n_1+..+n_{t-5}]}\; [1+\kappa n_{t-4}]^{-1}\;
\sum_{n_{t-3}=0}^{M-[n_1+..+n_{t-4}]}\; [1+\kappa n_{t-3}]^{-1}\;[1+\kappa (M-[n_1+..+n_{t-3}])]^{-p_2}
\nonumber
\ea
Comparing (\ref{2.30}) and (\ref{2.34}) we see that we can now iterate and define a sequence 
$p_u=2p_{u-1}-1,\; u=1,..,t-1$ of powers provided we can keep them in the range $0< p_u <1$. The iteration 
is easily solved by 
\be \label{2.35}
p_u=2^{u-1} p_1-(2^{u-1}-1)
\ee
Pick $p_1=1-2^{-T}$ then  
\be \label{2.36}
p_u=1-2^{u-1-T}
\ee
which satisfies $0<p_u<1$ for $u=1,..,t-1$ if $T\ge t$. As we have to perform $t-1$ summations 
in $s_t(M)$, for the choice $T=t$ we have $p_{t-1}=3/4$ and find 
that for some constant $c_t$ we have 
\be \label{2.37}
s_t(M)\le c_t\; [1+\kappa M]^{-3/4}
\ee
Correspondingly 
\be \label{2.38}
S_r(K)\le m^r\; [\sum_{t=1}^{r-1}\;
\left( \begin{array}{c} r \\ t \end{array} \right)\; c_t\; c_{r-t}]\;\;
\sum_{L=|K|}^\infty\; ([1+\kappa L]\;[1+\kappa(L-|K|)])^{-3/4}
\ee
Let $\kappa'={\sf min}(1,\kappa)$ so that $1+\kappa M\ge \kappa'(1+M)$ then for some 
constant $C_{r,m}$
\be \label{2.39}
S_r(K)\le C_{r,m} \;
\sum_{L=|K|}^\infty\; ([1+L]\;[1+L-|K|])^{-3/4}
\le C_{r,m}\; \zeta(\frac{3}{2})
\ee
where we introduced $L-|K|=0,1,2,..$ as a new summation variable and estimated 
$1+L+|K|\ge 1+L$ so that the Riemann zeta $\zeta(z)$ appears which converges for 
$\Re(z)>1$ (this fact can of course also proved elementarily using the above 
technique to estimate the sum by an integral). 
Note that the bound (\ref{2.39}) is in fact independent of $K$ and can certainly 
be optimised if needed.

\section{Renormalisation tools}
\label{sa}

More details on this section, in particular the relation to wavelet theory \cite{m},  can be found in \cite{l}.\\
\\
We work on spacetimes diffeomorphic to $\mathbb{R}\times\sigma$. In a first step the 
spatial D-manifold $\sigma$ is compactified to $T^D$. Therefore, all constructions 
that follow have to be done direction wise for each copy of $S^1$. On $X:=S^1$, understood 
as $[0,1)$ with endpoints identified, we consider 
the Hilbert space $L=L_2([0,1),\;dx)$ with orthonormal basis 
\be \label{a.1}
e_n(x):= e^{2\pi\;i\;n\;x},\; n\in\mathbb{Z}
\ee
with respect to the inner product 
\be \label{a.2}
<F,G>_L:=\int_0^1\;dx\; \overline{F(x)}\; G(x)
\ee 
Let $\mathbb{O}\subset\mathbb{N}$ be the set of positive odd integers. We equip $\mathbb{O}$ 
with a partial order, namely 
\be \label{a.3}
M<M' \;\;\Leftrightarrow\;\; \frac{M'}{M}\in \mathbb{N}
\ee
Note that this is not a linear order, i.e. not all elements of $\mathbb{O}$ are in relation,
but $\mathbb{O}$ is directed, that is, for each $M,M'\in \mathbb{O}$ we find 
$M^{\prime\prime}\in \mathbb{O}$ such that $M,M'<M^{\prime\prime}$ e.g. $M^{\prime\prime}=M M'$.  
For each $M\in \mathbb{O}$, called a resolution scale, we introduce the subsets 
$\mathbb{N}_M\subset \mathbb{N}_0,\;
\mathbb{Z}_M\subset \mathbb{Z},
X_M\subset X$ of respective cardinality $M$ defined by 
\be \label{a.4}
\mathbb{N}_M=\{0,1,..,M-1\},\; 
\mathbb{Z}_M=\{-\frac{M-1}{2}, -\frac{M-1}{2}+1,..,\frac{M-1}{2}\},\;
X_M=\{x^M_m:=\frac{m}{M},\; m\in \mathbb{N}_M\}        
\ee
It is easy to check that we have the lattice relation 
\be \label{a.5}
X_M\subset X_{M'} \;\;\Leftrightarrow\;\; M<M'
\ee
The subspace $L_M\subset L$ is defined by 
\be \label{a.6}
L_M:={\sf span}(\{e_n,\; n\in \mathbb{Z}_M\})
\ee
On $L_M$ we use the same inner product as on $L$, hence the $e_n,\; n\in \mathbb{Z}_M$ 
provide an ONB for $L_M$. An alternative basis for $L_M$ is defined by the functions
\be \label{a.7}
\chi^M_m(x):=\sum_{n\in \mathbb{Z}_M}\; e_n(x-x^M_m)
\ee
The motivation to introduce these functions is that in contrast to the plane waves 
$e_n$ they are 1. spatially concentrated at $x=x^M_m$ and 2. real valued. This 
makes them useful for renormalisation purposes. In addition, in contrast to 
characteristic functions which have better spatial location properties, they 
are smooth. This is a crucial feature because quantum field theory involves 
products of derivatives of the fields and derivatives of characteristic functions 
yield $\delta$ distributions. More in general, renormalisation tools must make 
a compromise between localisation and smoothness.      

The functions $\chi^M_m$ are still orthogonal but not orthonormal
\be \label{a.8}
<\chi^M_m,\;\chi^M_{\hat{m}}>_{L_M}=M\; \delta_{m,\hat{m}}
\ee
We choose not to normalise them in order to minimise the notational clutter in what 
follows. Let $l_M$ be the space of square summable sequences $f_M=(f_{M,m})_{m\in \mathbb{N}_M}$
with $M$ members and 
inner product given by 
\be \label{a.9}
<f_M,\;g_M>_{l_M}:=\frac{1}{M}\sum_{m\in\mathbb{N}_M}\; \overline{f_{M,m}}\; g_{M,m}  
\ee
If we interpret $f_{M,m}=F(x^M_m)$ then (\ref{a.9}) is a lattice approximant of 
$<F,G>_L$. We define 
\be \label{a.10}
I_M:\; l_M\to L_M;\; (I_M\cdot f_M)(x):=
<\chi^M_{\cdot}(x), f_M>_{l_M}=\frac{1}{M}\;\sum_{m\in \mathbb{N}_M}\; f_{M,m}\; \chi^M_m(x)
\ee
Its adjoint is defined by the requirement that 
\be \label{a.11}
<I_M^\ast\cdot F_M,\; g_M>_{l_M}= <F_M,\; I_M\cdot g_M>_{L_M}
\ee
which demonstrates 
\be \label{a.12}
I_M^\ast:\; L_M\to l_m,\; (I_M^\ast\cdot F_M)_m=<\chi^M_m, F_M>_{L_M}
\ee
One easily checks, using (\ref{a.8}) that 
\be \label{a.13}
I_M^\ast \cdot I_M=1_{l_M}, \;\;<I_M .,I_M .>_{L_M}=<.,.>_{l_M}
\ee
which shows that $L_M, l_M$ are in 1-1 correspondence and that $I_M$ is an isometry.
Likewise 
\be \label{a.13a}
P_M:=I_M \cdot I_M^\ast=1_{L_M}
\ee
We can consider $I_M$ also as a map $I_M:\; l_M\to L$ with image $L_M$ and then 
$<I_M^\ast ., .>_{l_M}=<.,I_M>_L$ shows that $I_M^\ast: L\to l_M$ is given by 
the same formula (\ref{a.12}) with $F\in L$ but now $P_M: L\to L_M$ is an orthogonal 
projection 
\be \label{a.14}
P_M\cdot P_M=P_M,\; P_M^\ast=P_M
\ee
We have explicitly 
\be \label{a.15}
(P_M\cdot F)(x)=\int_0^1\; dy\; P_M(x,y)\; F(y),\;
P_M(x,y)=\sum_{n\in \mathbb{Z}_M}\; e_n(x-y)
\ee
i.e. $P_M(x,y)$ is the M-cutoff of the $\delta$ distribution on $X$, i.e. modes  
$|n|>\frac{M-1}{2}$ are discarded. 

Given a continuum function 
$F\in L$ we call $f_M=I_M^\ast\cdot F\in l_M$ or $F_M=P_M\cdot F\in L_M$ the discretisation of 
$F$ at resolution 
$M$. In particular, if we have a Hamiltonian field theory on $X$ with conjugate pair of fields
$(\Phi,\Pi)$ i.e. the non-vanishing Poisson brackets are 
\be \label{a.16}
\{\Pi(x),\Phi(y)\}=\delta_X(x,y)=\sum_{n\in \mathbb{Z}}\; e_n(x-y)
\ee
then their discretisations obey
\be \label{a.16a}
\{\pi_{M,m},\phi_{M,\hat{m}}\}=M\;\delta_{m,\hat{m}},\;\;
\{\Pi_M(x),\Phi_M(y)\}=P_M(x,y)=\frac{\sin(M\pi(x-y))}{\sin(\pi(x-y))}
\ee
The latter formula is known as the Dirichlet kernel.

Given a functional $H[\Pi,\Phi]$ of the continuum fields we define its discretisation 
by 
\be \label{a.16b}
h_M[\pi_M,\phi_M]:=H_M[\Pi_M,\Phi_M]=H[\Pi_M,\Phi_M]=(I'_M H)[\pi_M,\phi_M]
\ee
where $I'_M$ denotes the pull-back by $I_M$. That is,
in the continuum formula for $H$ one substitutes $\Pi\to \Pi_M,\; \Phi\to \Phi_M$
in the formula for $H$ 
upon which $H$ is restricted to $\Pi_M,\Phi_M$, i.e. $H_M$ is that restriction, 
and then uses the identity $\Pi_M=I_M\cdot \pi_M,\;\Phi_M=I_M\cdot \phi_M$. In order for this 
to be well-defined it is important that $I_M$ is sufficiently smooth as $H$ typically 
depends of derivatives of $\Pi,\Phi$. This is granted by our choice of $I_M$. In particular,
as the derivative $\partial=\frac{\partial}{\partial x}$ preserves each of the spaces $L_M$ 
we have a canonical discretisation of the derivative defined by 
\be \label{a.17}
\partial_M:=I_M^\ast \cdot \partial\cdot I_M
\ee
which obeys $\partial_M^n=I_M^\ast \cdot \partial^n\cdot I_M$ because 
$I_M\cdot I_M^\ast=P_M$ and $[\partial,P_M]=0$. 

Concerning quantisation, in the the continuum we define the Weyl algebra $\mathfrak{A}$ 
generated by the Weyl elements 
\be \label{a.18}
W[F]=e^{-i<F,\Phi>_L},\; W[G]=e^{-i<G,\Pi>_L}
\ee 
for real valued $F,G\in L$ (or a dense subspace thereof with additional properties 
such as smoothness and rapid momentum decrease of its Fourier modes 
$<e_n,F>_L, \; <e_n,G>_L$). That is, the non-trivial Weyl relations are  
\ba \label{a.19}
&& W[G]\; W[F]\; W[-G]=e^{-i<G,F>_L}\;W[F],\; 
W[F]\;W[F']=W[F+F'], \; W[G]\;W[G']=W[G+G']
\nonumber\\
&& W[0]=1_{\mathfrak{A}},\;
W[F]^\ast=W[-F],\; W[G]^\ast=W[-G] 
\ea
Cyclic representations $(\rho,{\cal H},\Omega)$ of $\mathfrak{A}$ with $\Omega\in {\cal H}$ 
a cyclic vector (i.e. ${\cal D}:=\rho(\mathfrak{A}){\cal H}$ is dense) are generated from states 
(positive, linear, normalised functionals)
$\omega$ on $\mathfrak{A}$ via the GNS construction \cite{p}. The correspondence is given 
by 
\be \label{a.20}
\omega(A)=<\Omega, \; \rho(A)\Omega>_{{\cal H}}  
\ee

We may proceed analogously with the discretised objects. For each $M$ we define 
the Weyl algebra $\mathfrak{A}_M$ generated by the Weyl elements  
\be \label{a.21}
W_M[F_M]=e^{-i<F_M,\Phi_M>_{L_M}}=w_M[f_M]=e^{-i<f_M,\phi_M>_{l_M}},\;\;
W_M[G_M]=e^{-i<G_M,\Pi_M>_{L_M}}=w_M[g_M]=e^{-i<g_M,\pi_M>_{l_M}}
\ee
where $F_M=I_M \cdot f_M,\; G_M=I_M\cdot g_M$ are real valued. Accordingly
\ba \label{a.22}
&&W_M[G_M]\; W_M[F_M]\; W_M[-G_M]=e^{-i<G_M,F_M>_{L_M}}\;W_M[F_M],\; 
W_M[F_M]\; W_M[F'_M]=W_M[F_M+F'_M],\;
\\
&& 
W_M[G_M]\; W_M[G'_M]=W_M[G_M+G'_M],\;
W_M[0]=1_{\mathfrak{A}_M},\;W_M[F_M]^\ast=W_M[-F_M],\; W_M[G_M]^\ast=W_M[-G_M] 
\nonumber
\ea 
and completely analogous for $\phi_M,\pi_M$ if we substitute lower case letters for capital letters
in (\ref{a.22}).
For each $M$ we define a state $\omega_M$ on $\mathfrak{A}_M$ which gives rise 
to GNS data $(\rho_M, {\cal H}_M, \Omega_M)$ and the dense subspace ${\cal D}_M=
\mathfrak{A}_M \Omega_M$. 
Note that $\mathfrak{A}_M$ is a subalgebra of $\mathfrak{A}_{M'}$ for $M<M'$ and 
that $\mathfrak{A}_M$ is a subalgebra of $\mathfrak{A}$. This follows from the 
identities 
\be \label{a.23}
W_{M'}[F_M]=W_M[F_M],\;\;W_M[F_M]=W[F_M]
\ee
due to $P_{M'} \cdot P_M= P_M$ since $L_M\subset L_{M'}$ and $P_M\cdot P_M=P_M$ respectively.
 
The sole reason for discretisation is as follows: While finding states on $\mathfrak{A}$ 
is not difficult (e.g. Fock states) it is tremendously difficult to find such states 
which allow to define non-linear functionals of $\Pi,\Phi$ such as Hamiltonians 
densely on $\cal D$ due to UV singularities arising from the fact that $\Pi,\Phi$ 
are promoted to operator valued distributions whose product is a priori ill-defined.
In the presence of the UV cut-off $M$ this problem can be solved because e.g. 
$\Phi_M(x)^2$ is perfectly well-defined ($\Phi$ is smeared with the smooth kernel 
$P_M$). Suppose then that $h_M$ or equivalently $H_M$ are somehow quantised
on ${\cal D}_M$. We denote these quantisations by 
$\rho_M(h_M,c_M)$ or $\rho_M(H_M,c_M)$ respectively to emphasise that these operators are 
1. densely defined on $\rho_M(\mathfrak{A}_M)\Omega_M$, 2. correspond to the classical symbol 
$h_M$ of $H_M$ respectively and 3. depend on a set of choices $c_M$ for each $M$ 
such as  factor or normal ordering etc. It is therefore not at all clear 
whether the theories defined for each $M$ in fact descend from a continuum theory. 
By ``descendance'' we mean that $\omega_M$ is the restriction of $\omega$ to 
$\mathfrak{A}_M$ and that $\rho_M(H_M,c_M)$ is the restriction of $\rho(H,c)$ to ${\cal D}_M$
as a quadratic form (i.e. in the sense of matrix elements). In formulas this means 
\begin{align} \label{a.24}
\omega_M(A_M)&=\omega(A_M), \\
<\rho_M(A_M)\Omega_M,\;\rho_M(H_M,c_M)\; \rho_M(B_M)\Omega_M>_{{\cal H}_M}
&=<\rho(A_M)\Omega,\;\rho(H,c)\; \rho(B_M)\Omega>_{{\cal H}}, \nonumber
\end{align}
for all $M\in \mathbb{O}$ and all $A_M,B_M\in\mathfrak{A}_M$. If they did,
then we obtain the following identities for $M<M'$
\begin{align} 
\label{a.25}
\omega_{M'}(A_M) & =\omega_M(A_M),\\
<\rho_M(A_M)\Omega_M,\;\rho_M(H_M,c_M)\; \rho_M(B_M)\Omega_M>_{{\cal H}_M}
& = <\rho_{M'}(A_M)\Omega_{M'},\;\rho_{M'}(H_{M'},c_{M'})\; \rho_{M'}(B_M)\Omega_{M'}>_{{\cal H}_{M'}}, \nonumber
\end{align}

called consistency conditions. This follows from the fact that $A_{M'}:=A_M,
B_{M'}:=B_M$ can be considered as elements of $\mathfrak{A}_M$ and then using (\ref{a.24}).
With some additional work \cite{d} one can show that (\ref{a.25}) are necessary and sufficient
for $\omega, \rho(H)$ to exist (at least as a quadratic form).

In constructive quantum field theory (CQFT) \cite{b} one proceeds as follows. One starts with 
an Ansatz of a family of discretised theories
$(\omega^{(0)}_M,\rho^{(0)}_M(H_M,c^{(0)}_M))_{M\in \mathbb{O}}$. That Ansatz generically 
violates (\ref{a.25}). We now define a renormalisation flow of states and quantisations 
by defining the sequence   
$(\omega^{(k)}_M,\rho^{(k)}_M(H_M,c^{(k)}_M))_{M\in \mathbb{O}}$ for $k\in \mathbb{N}_0$ via 
\ba\label{a.26}
&& \omega^{(k+1)}_M(A_M):=\omega^{(n)}_{M'(M)}(A_M),\;\;
<\rho_M^{(k+1)}(A_M)\Omega^{(k+1)}_M,\;\rho^{(k+1)}_M(H_M,c^{(k+1)}_M)\; \rho^{(k+1)}_M(B_M)\Omega^{(k+1)}_M>_{{\cal H}^{(k+1)}_M}
\nonumber\\
&=& <\rho^{(k)}_{M'}(A_M)\Omega^{(k)}_{M'},\;\rho^{(k)}_{M'}(H_{M'},c^{(k)}_{M'})\; \rho^{(k)}_{M'}(B_M)\Omega^{(k)}_{M'}>_{{\cal H}^{(k)}_{M'}}
\ea
where $M':\mathbb{O}\to \mathbb{O}$ is a fixed map with the property that 
$M'(M)>M,\;M'(M)\not=M$. The first relation defines a new state at the coarser 
resolution $M$ as the restriction of the old state at the finer resolution $M'(M)$. 
This then defines also new GNS data $(\rho^{(k+1)}_M,{\cal H}^{(k+1)}_M,\Omega^{(k+1)}_M)$ via the GNS construction.
The second relation defines the matrix elements of an operator or quadratic form
in that new representation 
and with new quantisation choices to be made at coarser 
resolution 
in terms of the restriction of the matrix elements of the old operator or quadratic form with old 
quantisation choices in the old representation at finer resolution. A fixed point 
family $(\omega^\ast_M,\rho^\ast_M(H_M,c^\ast_M))_{M\in \mathbb{O}}$
of the flow (\ref{a.26}) solves (\ref{a.25}) at least for $M'=M'(M)$ and all $M$
and thus all $[M']^n(M),\;n\in \mathbb{N}_0$ and all $M$. This typically implies 
that (\ref{a.25}) holds for all $M'<M$. In practice we will work with $M'(M):=3
\;M$  
  
Note that for a general operator or quadratic form $O$ defined densely on $\cal D$
it is not true that we find an element $a\in \mathfrak{A}$ such that $\rho(a)=O$
(e.g. unbounded operators) which is why the above statements cannot be made just in terms of 
the states $\omega$. If one tried, one would need to use sequences or nets 
$a_n\in \mathfrak{A}$ whose
limits lie outside of $\mathfrak{A}$. On the other hand, if one prefers to work with the Weyl elements
$W_M[F_M]$ one may relate the spaces $l_M, \; L_M$ via the identities 
$w_M[f_M]=W_M[F_M],\; f_M=I_M^\ast\cdot F_M$. The $w_M[f_M], \; w_{M'}[f_M']$ at resolution $M, \; M'$ respectively can be related via the \textit{coarse graining} map $I_{M M'}:= I_{M'}^\ast\cdot I_M;\; l_M\to l_{M'}$ such that
$w_{M'}[I_{M M'}\cdot f_M]=w_M[f_M]$. This map obeys $I_{M_2 M_3}\cdot I_{M_1 M_2}=I_{M_1 M_3}$ for 
$M_1<M_2<M_3$ because the image of $I_M$ is $L_M$ which is a subspace of $L_{M'}$ thus
$I_{M_2 M_3}\cdot I_{M_1 M_2}=I_{M_3}^\ast\cdot P_{M_2}\cdot  I_{M_1}=I_{M_3}^\ast\cdot I_{M_1}$. 
Then  $W_{M'}[F_M]=w_{M'}[I_{M'}^\ast F_M]=w_{M'}[I_{M'}^\ast\cdot I_M\cdot f_M]
=w_{M'}[I_{M M'}\cdot f_M]$ indeed. For the same reason $W_{M'}[F_M]=W_M[F_M]$ as 
$L_M$ is embedded in $L_{M'}$ by the identity map. The renormalisation flow in terms of Weyl elements $w_M[f_M]$ and the coarse graining map $I_{M,M'}$ takes the form
\begin{align}
\label{rflow}
\omega_M^{(k+1)}(w_M[f_M])&:=\omega^{(k)}_{M'}(w_{M'}[I_{M,M'}f_M']) \nonumber \\ 
\braket{w_M[f'_M]\Omega_{M}^{(k+1)},H^{(k+1)}_{M}\;w_M[f_M]\Omega_{M}^{(k+1)}}_{\mathcal{H}^{(k+1)}_M}&:=\braket{w_{M'}[I_{M,M'}f'_M]\Omega_{M'}^{(k)},H^{(k)}_{M'}\;w_{M'}[I_{M,M'}f_M]\Omega_{M'}^{(k)}}_{\mathcal{H}^{(k)}_{M'}}.
\end{align}

\end{appendix}

}


\begin{thebibliography}{99}

\parskip -5pt

\bibitem{a} R. Haag, ``Local Quantum Physics'', Springer Verlag, Berlin,
1984

\bibitem{b} J. Glimm and A. Jaffe, ``Quantum Physics'',
Springer Verlag, New York, 1987.

\bibitem{c} K. G. Wilson. The renormalization group:
Critical phenomena and the Kondo
problem. Rev. Mod. Phys. {\bf 47} (1975) 773

\bibitem{d} T. Thiemann. Canonical quantum gravity, constructive QFT and
renormalisation.
Front. in Phys. \textbf{ 8} (2020) 548232. arXiv:2003.13622 [gr-qc].

\bibitem{LLT1} T. Lang, K. Liegener, T. Thiemann.
 Hamiltonian Renormalisation I.
Derivation from Osterwalder-Schrader Reconstruction.
Class. Quant. Grav. {\bf 35} (2018) 245011.
[arXiv:1711.05685]

\bibitem{LLT2} T. Lang, K. Liegener, T. Thiemann. Hamiltonian Renormalisation II.
Renormalisation Flow of 1+1 dimensional free, scalar fields: Derivation.
Class. Quant. Grav. {\bf 35} (2018) 245012.
[arXiv:1711.06727]

\bibitem{LLT3} T. Lang, K. Liegener, T. Thiemann. Hamiltonian Renormalisation III.
Renormalisation Flow of 1+1 dimensional free, scalar fields: Properties.
Class. Quant. Grav. {\bf 35} (2018) 245013.
[arXiv:1711.05688]

\bibitem{LLT4} T. Lang, K. Liegener, T. Thiemann. Hamiltonian Renormalisation IV. Renormalisation Flow of D+1 dimensional
free scalar fields and Rotation Invariance.
Class. Quant. Grav. {\bf 35} (2018) 245014, [arXiv:1711.05695]

\bibitem{LT} K. Liegener, T. Thiemann.
Hamiltonian Renormalisation V. Free Vector Bosons.
Front. Astron. Space Sci. {\bf 7} (2021) 547550.
e-Print: 2003.13059 [gr-qc]

\bibitem{TT} T. Thiemann.
Hamiltonian Renormalisation VII. Free Fermions and doubler free kernels.
Phys. Rev. {\bf D108} (2023) 12, 125007.
e-Print: 2207.08291 [hep-th]

\bibitem{TZ} E.-A. Zwicknagel. Hamiltonian renormalization. VI. Parametrized field theory on the cylinder.
Phys. Rev. {\bf D108} (2023) 12, 125006. e-Print: 2207.08290 [gr-qc]

\bibitem{e} B. Simon, ``The P($\phi$)2 Euclidean (Quantum) Field Theory'',
Princeton Unviersity Press, 1974

\bibitem{m} I. Daubechies. Ten lectures of wavelets. Springer Verlag, Berlin,
1993.\\
A. Cohen, I. Daubechies, P.Vial. Wavelets on the interval and
fast wavelet transforms. Appl. and Comp. Harm. Anlysis, Elsevier, 1993.
[hal-01311753]

\bibitem{l} T. Thiemann. Renormalization, wavelets, and the Dirichlet-Shannon kernels. Phys. Rev. D {\bf 108} 
(2023) 12, 125008. e-Print: 2207.08294 [hep-th]

\bibitem{p} O. Bratteli, D. W. Robinson, ``Operator Algebras and Quantum
Statistical
Mechanics'', vol. 1,2, Springer Verlag, Berlin, 1997.

\bibitem{k} J. Glimm, ``Boson Fields with the $:\phi^4:$ Interaction in Three
Dimensions", Comm. Math. Phys. {\bf 10} (1968)  1-47.\\
J. Glimm, A. Jaffe, Positivity of the $\phi^4_3$ Hamiltonian'',
Fortschr. Phys. {\bf 21} (1973) 327–376

\bibitem{f}  L. Smolin. The G(Newton) $\to 0$
limit of Euclidean quantum gravity. Class. Quant. Grav. {\bf 9} (1992)
883-894. e-Print: hep-th/9202076 [hep-th]

\bibitem{g} Exact quantisation of U(1)$^3$ quantum gravity via exponentiation of the hypersurface deformation algebroid.
Class. Quant. Grav. {\bf 40} (2023) 24, 245003. e-Print: 2207.08302 [gr-qc]

\bibitem{h} H. Narnhofer, W.E. Thirring.
Covariant QED without indefinite metric.
Rev. Math. Phys. {\bf 4} (1992) spec01, 197-211

\bibitem{i} T. Thiemann. Nonperturbative quantum gravity in Fock representations.
Phys. Rev. {\bf D 110} (2024) 12, 124023. e-Print: 2405.01212 [gr-qc]

\bibitem{j} C. Rovelli, ``Quantum Gravity'', Cambridge University
Press, Cambridge, 2004.\\
T. Thiemann, ``Modern Canonical Quantum General Relativity'', Cambridge
University Press, Cambridge, 2007.\\
J. Pullin, R. Gambini, ``A first course in Loop Quantum Gravity'',
Oxford University Press, New York, 2011\\
C. Rovelli, F. Vidotto, ``Covariant Loop Quantum Gravity'', Cambridge
University Press, Cambridge, 2015\\
K. Giesel, H. Sahlmann,
From Classical To Quantum Gravity: Introduction to Loop Quantum Gravity,
PoS QGQGS2011 (2011) 002, [arXiv:1203.2733].

\bibitem{OS} K. Osterwalder, R. Schrader. Axioms for Euclidean Green's functions. Communications in Mathematical Physics \textbf{31}(2):83–112, 6 1973

\bibitem{nocutoffs} 
J. Glimm and A. Jaffe, "A $\lambda\phi^4$ quantum field theory without cutoffs. I", Phys. Rev. \textbf{176} (1968), 1945–1951. DOI: 10.1103/PhysRev.176.1945. \\
J. Glimm and A. Jaffe, "The $\lambda(\phi^4)_2$ quantum field theory without cutoffs: III. The physical vacuum", Acta Math. \textbf{125} (1970), 203–267. DOI: 10.1007/BF02392335. \\
J. Glimm and A. Jaffe, "The $\lambda(\phi^4)_2$ quantum field theory without cutoffs. II. The field operators and the approximate vacuum", Ann. of Math. \textbf{91} (1970), no. 2, 362–401. DOI: 10.2307/1970582. \\
J. Glimm and A. Jaffe, "The $\lambda(\phi^4)_2$ quantum field theory without cutoffs. IV. Perturbations of the Hamiltonian", J. Math. Phys. \textbf{13} (1972), no. 10, 1568–1584. DOI: 10.1063/1.1665879.

\end{thebibliography}
\end{document}